\documentclass[a4paper, amsfonts, amssymb, amsmath, reprint, showkeys, nofootinbib, twoside,notitlepage,onecolumn]{revtex4-1}

\def\jqsrt{{J.~Quant.~Spec.~Radiat.~Transf.}}

\usepackage{amsmath,amstext}
\usepackage[T1]{fontenc}
\usepackage{amssymb}
\usepackage{graphicx}
\usepackage{ae,aecompl}

\DeclareFontFamily{OT1}{pzc}{}
\DeclareFontShape{OT1}{pzc}{m}{it}{<-> s * [1.10] pzcmi7t}{}
\DeclareMathAlphabet{\mathpzc}{OT1}{pzc}{m}{it}

\usepackage{hyperref}
\usepackage{amsmath}
\usepackage{amssymb}
\usepackage{mathtools}
\usepackage{bm}
\usepackage{cleveref}
\usepackage{tensor}
\usepackage{braket}
\usepackage{enumitem}
\usepackage{mhchem}
\usepackage{amsthm}
\usepackage{nccmath}
\usepackage{mathrsfs}
\usepackage{color}

\newcommand{\n}{{\mathbf{n}}}

\newcommand{\PN}{{\mathbb{P}_N}}
\newcommand{\s}{{\sigma \!\! \! \sigma}}

\def\be{\begin{equation}}
\def\ee{\end{equation}}
\def\beq{\begin{eqnarray}}
\def\eeq{\end{eqnarray}}

\theoremstyle{definition}

\theoremstyle{theorem}

\theoremstyle{corollary}

\begin{document}
\title{Exact resummation of hydrodynamics for arbitrary elastic scattering integral}
\author{L.~Gavassino}
\affiliation{Department of Applied Mathematics and Theoretical Physics, University of Cambridge, Wilberforce Road, Cambridge CB3 0WA, United Kingdom}

\begin{abstract}
We consider monoenergetic particles propagating in three spatial dimensions and undergoing elastic scattering off an external medium. For an arbitrary rotationally invariant scattering kernel, we derive an exact expression for the frequency- and wave-number-dependent diffusivity, thereby resumming the hydrodynamic gradient expansion to all orders. The result is an infinite continued fraction, with each successive level determined by the relaxation rate of a higher angular harmonic of the collision operator. Thus, while ordinary diffusion probes only the first angular harmonic, the resummed constitutive relation retains information about the full scattering kernel. We show how this microscopic information manifests itself in the analytic structure of the diffusivity and in the response to a localized external source. In particular, deviations from Fick's law decay exponentially with distance, at a rate determined by the nearest complex-wavenumber singularity of the inverse diffusivity. Thus, exponentially small departures from hydrodynamics retain detailed information about the underlying scattering process.
\end{abstract} 
\maketitle

\vspace{-0.9cm}
\section{Introduction}
\vspace{-0.3cm}

Hydrodynamic constitutive relations are usually organized as expansions in spacetime gradients \cite{Baier2008,Struchtrup2011Review,RomatschkeReview:2017ejr,FlorkowskiReview2018,Kovtun2019,BemficaDNDefinitivo2020,GavassinoInfiniteOrded2024pgl}. Beyond leading order, these corrections account for the fact that dissipative fluxes do not respond instantaneously and locally to variations of the conserved densities: microscopic dynamics introduces delays both in time and in space. Describing such nonlocal response through a local gradient expansion requires, in general, infinitely many transport coefficients. When the expansion converges, however, this infinite hierarchy can be resummed into a single spectral function, which may be interpreted as a frequency- and wavenumber-dependent transport coefficient \cite{ChungYip1969,Hess1983,TGaskell_1987,VelascoColin,Hansen2007,AlfordBulk,StephanovHydroPlus:2017ghc,Kiselev,BulkGavassino,Krihin2023,Nazaryan2024}.

In a companion paper \cite{GavassinoDiamond:2026fgz}, we have developed a rigorous technique for computing the resummed transport coefficients in a broad class of causal kinetic-type theories. For a single conserved particle density \(\rho\), diffusing according to the continuity equation (assuming the gradient point in the $x$ direction)
\begin{equation}\label{continuity}
\partial_t \rho+\partial_x J=\text{``External source''} \, ,
\end{equation}
we showed that there always exists, in the linear regime, a Fourier-transformed constitutive relation of the form
\begin{equation}\label{constitutiverelation}
J=-ik\,\mathfrak{D}(i\omega,ik)\,\rho ,
\end{equation}
where the resummed diffusivity \(\mathfrak{D}(i\omega,ik)\) is the relevant spectral function of the problem. Its Taylor coefficients around \(\omega=k=0\) are all the transport coefficients of the gradient expansion, while the full function captures the spatial and temporal nonlocality of the response. For causal theories with maximal propagation speed \(w\) and nonhydrodynamic gap \(1/\tau\), we showed that this expansion converges at least within the domain $|i\omega|+w|ik|<1/\tau.$

This raises the question of whether the resummed diffusivity can be calculated directly from microscopic dynamics in a fully realistic kinetic model, with an arbitrarily complicated scattering integral, without constructing the gradient expansion order by order.
We answer this question positively for the transport of monoenergetic particles, such as photons or neutrons, propagating in three spatial dimensions with fixed speed \(w\). The particles do not interact with one another, but undergo elastic scattering off a homogeneous background medium. Our only assumption will be rotational invariance of the background medium. At long time and length scales, repeated scattering gives rise to ordinary diffusion. Beyond this limit, the finite propagation time between collisions generates an infinite hierarchy of gradient corrections, whose structure depends on the angular properties of the microscopic scattering process.

We will show that this hierarchy can be resummed exactly. For an arbitrary rotationally invariant elastic scattering kernel, we obtain the full diffusivity \(\mathfrak{D}(i\omega,ik)\) analytically in the form of an infinite continued fraction determined by the angular relaxation rates of the collision operator. The result requires neither a relaxation-time nor a small-angle approximation and applies throughout the convergence domain of the hydrodynamic gradient expansion. It therefore provides an explicit example in which an infinite set of transport coefficients can be resummed exactly, while retaining full information about the underlying scattering kernel.

Throughout the article, we adopt the metric signature $(-,+,+,+)$, and work in natural units, $c=k_B=\hbar=1$.

\vspace{-0.3cm}
\section{Derivation of the resummed diffusivity}\label{resummeddiffusivitygneeraltheory}
\vspace{-0.3cm}

In this section, we derive the exact expression for the resummed diffusivity $\mathfrak{D}(i\omega,ik)$ governing the transport of monoenergetic particles with speed $w$, propagating in three spatial dimensions, and undergoing elastic scattering off a homogeneous, rotationally invariant background medium. The derivation is based on \cite{GavassinoDiamond:2026fgz}.

\subsection{Transport equation with isotropic sources}

Let \(I(x^\mu,\mathbf n)\) be the angular distribution function, which counts how many particles occupy the spacetime point \(x^\mu\) and propagate in direction \(\mathbf n\in S^2\). We regard it as a function \(I:\text{``Spacetime''}\to L^2(S^2)\), so that, at a given event, we have the local inner product \((I,\widetilde{I})=\int I^*(\mathbf n)\widetilde{I}(\mathbf n)d^2\Omega\). We choose the normalization of \(I\) such that the total particle density is \(\rho=(Y_0^0,I)\), with \(Y_0^0=1/\sqrt{4\pi}\) the normalized isotropic spherical harmonic.

Along configurations with gradients pointing in direction \(x\), the Boltzmann equation reads \cite{case1967linear,pomraning1973equations,Levermore1984,mihalas_book}
\begin{equation}\label{boltzmann}
(\partial_t +wn^x \partial_x)I =-\s I +Q(t,x) Y_0^0 \, ,
\end{equation}
where \(\s\) is the linearized scattering operator and \(QY_0^0\) is an external source, which we assume to be isotropic. The operator \(\s\) is self-adjoint in \(L^2(S^2)\) by Onsager symmetry, and non-negative by the second law \cite{DudynskiEkielJezewska1985,GavassinoSymmetric2022nff,GavassinoDisturbing:2026klp}. We further assume that its nonhydrodynamic spectrum is gapped \cite{Dudynski1989Hydro,GavassinoGapless:2024rck,GavassinoMemoryEffects:2026ebx}, namely \(\text{Spectrum}(\s)\subseteq \{0\}\cup [1/\tau,\infty)\), with \(\tau\) the longest relaxation time, and that \(\ker(\s)=\text{span}(Y_0^0)\), so that particle number is the only conserved quantity. Indeed, taking the inner product of \eqref{boltzmann} with \(Y_0^0\), we recover the continuity equation \eqref{continuity}, with \(Q\) as the external source, and
\begin{equation}\label{theflux}
J=w(Y_0^0,n^x I) \, .
\end{equation}

\subsection{Resummed diffusivity as a matrix element}

Suppose that \(Q\propto e^{ikx-i\omega t}\). We may then seek a response \(I\) with the same spacetime dependence, so that \eqref{boltzmann} reduces to \((\s+ikw n^x-i\omega)I=QY_0^0\). Applying on both sides the orthogonal projector \(\PN=1-Y_0^0(Y_0^0,\cdot)\), and introducing the orthogonal decomposition \(I=(1-\PN)I+\PN I=\rho Y_0^0+I_N\), we obtain
\begin{equation}
(\s_N+ikw n^x_N-i\omega) I_N =-ikw\PN n^x Y_0^0 \rho \, ,
\end{equation}
where we have introduced the compressed operators \(\s_N=\PN\s \PN\) and \(n_N^x=\PN n^x \PN\), both viewed as operators on \((Y_0^0)^\perp\). Since \((Y_0^0,n^xY_0^0)=0\), we also have \(\PN n^xY_0^0=n^xY_0^0\).

For \(|i\omega|+w|ik|<1/\tau\), the operator \(\s_N+ikw n^x_N-i\omega\) is invertible. Hence, we can solve for the nonhydrodynamic part of the distribution as \(I_N=-ikw[\s_N+ikw n^x_N-i\omega]^{-1}n^xY_0^0\rho\). Plugging this expression into \eqref{theflux}, we recover equation \eqref{constitutiverelation}, with
\begin{equation}\label{matrixelment}
\mathfrak{D}(i\omega,ik)=w^2(n^x Y_0^0, [\s_N+ikw n^x_N-i\omega]^{-1}n^x Y_0^0) \, ,
\end{equation}
which agrees with the general formula obtained in the companion paper. Our task is now to evaluate this matrix element explicitly for an arbitrary rotationally invariant scattering operator.

\subsection{Consequences of rotational invariance}

The requirement that $\s$ be invariant under rotations translates into the statement that $\s Y_\ell^m =\sigma_\ell Y_\ell^m$. Moreover, working in spherical coordinates with the polar axis along $x$, we can write $n^x=\cos\theta$, whose action on the spherical harmonics is
\begin{equation}
\cos\theta \, Y_\ell^0 =a_{\ell+1}Y^0_{\ell+1}+a_\ell Y^0_{\ell-1} \, , \qquad\qquad \text{with } a_\ell = \dfrac{\ell}{\sqrt{4\ell^2-1}} \, .
\end{equation}
Thus, the $m=0$ sector does not communicate with other $m$ sectors, and since the matrix element defining the resummed diffusivity is ultimately taken between $Y_0^0$, we can work entirely within such sector.

In the orthonormal basis $\{Y_1^0,Y_2^0,...\}$, we have
\begin{equation}\label{sNnN}
\s_N+ikw n^x_N-i\omega=
\begin{bmatrix}
\sigma_1{-}i\omega & ikwa_2 & 0 & 0 & 0 &\cdots\\
ikwa_2 & \sigma_2{-}i\omega & ikwa_3 & 0 & 0 &\cdots\\
0 & ikwa_3 & \sigma_3{-}i\omega & ikwa_4 & 0 &\cdots\\
0 & 0 & ikwa_4 & \sigma_4{-}i\omega & ikwa_5 & \cdots \\
0 & 0 & 0 & ikwa_5 & \sigma_5{-}i\omega &\cdots\\
\vdots & \vdots & \vdots & \vdots & \vdots & \ddots \\ 
\end{bmatrix} \, ,
\end{equation}
and we only need the first diagonal entry of its inverse, since
\begin{equation}\label{Dwa1}
\mathfrak{D}(i\omega,ik)=w^2 a_1^2(Y^0_1,[\s_N+ikw n^x_N-i\omega]^{-1}Y^0_1) \, .
\end{equation}
This matrix element can be evaluated using the standard block-matrix formula
\begin{equation}
\text{Upper-left block}
\begin{bmatrix}
A & B \\
C & D \\
\end{bmatrix}^{-1}
=(A-BD^{-1}C)^{-1} \, .
\end{equation}
In particular, let $\mathbb{M}^{(\ell)}$ denote the semi-infinite matrix obtained by restricting $\s_N+ikw n_N^x-i\omega$ to the sector spanned by \(\{Y_\ell^0,Y_{\ell+1}^0,\ldots\}\). Separating out its first row and column recursively, we obtain
\begin{equation}\label{intermediateM}
\begin{split}
[\s_N+ikw n^x_N-i\omega]^{-1}_{11} &\equiv[\mathbb{M}^{(1)}]^{-1}_{11} \\
&= \dfrac{1}{\sigma_1{-}i\omega-(ikwa_2)^2 [\mathbb{M}^{(2)}]^{-1}_{11} }  \\
&= \dfrac{1}{\sigma_1{-}i\omega-\dfrac{(ikwa_2)^2}{\sigma_2{-}i\omega -(ikwa_3)^2 [\mathbb{M}^{(3)}]^{-1}_{11}} }  \\
&= \dfrac{1}{\sigma_1{-}i\omega-\dfrac{(ikwa_2)^2}{\sigma_2{-}i\omega -\dfrac{(ikwa_3)^2}{\sigma_3{-}i\omega-(ikwa_4)^2[\mathbb{M}^{(4)}]^{-1}_{11}} } } =... \\
\end{split}
\end{equation}
Hence, the resummed diffusivity takes the form of a continued fraction:
\begin{equation}\label{Diomik}
\boxed{
\mathfrak{D}(i\omega,ik)= \dfrac{w^2 a_1^2}{\sigma_1{-}i\omega-\dfrac{(ikwa_2)^2}{\sigma_2{-}i\omega -\dfrac{(ikwa_3)^2}{\sigma_3{-}i\omega-\dfrac{(ikwa_4)^2}{\sigma_4{-}i\omega \ddots}} } } \, .}
\end{equation}
This is our main result (see also \cite{Bogumarian2000,Krihin2023,Nazaryan2024} for similar continued fractions in other setups). It shows that the full angular dependence of the microscopic scattering process, encoded in the sequence of relaxation rates $\{\sigma_\ell\}_{\ell\geq1}$, survives in the all-order hydrodynamic constitutive relation. In the remainder of this manuscript, we examine this function for physically relevant scattering integrals.

\subsection{General remarks}

In the long-wavelength, low-frequency limit, we recover the ordinary diffusion coefficient, $\mathfrak{D}(0,0)=w^2/(3\sigma_1)$, which agrees with the standard random-walk formula upon identifying $\sigma_1^{-1}$ with the transport relaxation time\footnote{In a random walk, $\mathfrak{D}=\int_0^\infty \langle v_x(t)v_x(0)\rangle dt$, with $v_x$ the velocity component in direction $x$ \cite[\S 9.1]{peliti_book}. Taking $\langle v_x^2\rangle=w^2/3$, by rotational invariance, and $\langle v_x(t)v_x(0)\rangle\propto e^{-\sigma_1t}$, we obtain $\mathfrak{D}=\int_0^\infty \frac{w^2}{3}e^{-\sigma_1t}dt=\frac{w^2}{3\sigma_1}$.}. Moreover, if we work at finite frequency but vanishing wavenumber, we obtain
\begin{equation}
\mathfrak{D}(i\omega,0)=\dfrac{\mathfrak{D}(0,0)}{1-i\omega\sigma_1^{-1}} \, ,
\end{equation}
which is indistinguishable from the resummed diffusivity of the Cattaneo equation, with relaxation time $\sigma_1^{-1}$.

Additionally, setting $\omega=-i\mathfrak{D}(0,0)k^2+\mathcal{O}(k^4)$ and expanding $\mathfrak{D}(i\omega,ik)$ to order $k^2$, we obtain
\begin{equation}
\mathfrak{D}_\text{on-shell}(k)=\dfrac{w^2}{3\sigma_1} + \dfrac{w^4}{9\sigma_1^3} \left[1-\dfrac{4\sigma_1}{5\sigma_2}\right]k^2+\mathcal{O}(k^4) \, .
\end{equation}
The coefficient of the $k^2$ correction determines the super-Burnett coefficient \cite{StruchtrupSuperBurnett2005}. More generally, every time we increase the order in $k^2$, an additional eigenvalue of $\s$ enters the hydrodynamic dispersion relation: the coefficient at order $k^{2a}$ depends on $\{\sigma_1,\ldots,\sigma_{a+1}\}$. Thus, progressively higher orders in hydrodynamics probe progressively finer angular structure of the microscopic scattering kernel.

\section{Two simple scattering models}

In this section, we analyze the resummed diffusivity for two limiting models: isotropic scattering and small-angle scattering, respectively,
\begin{equation}
\begin{split}
&\s = \tau^{-1} \PN \qquad \qquad \qquad\quad\, (\text{isotropic scattering}) \, , \\
&\s = -(2\tau)^{-1} \Delta_{LB} \qquad \qquad \, \,\, (\text{small-angle scattering}) \, ,
\end{split}
\end{equation}
where $\Delta_{LB}$ is the Laplace--Beltrami operator on $S^2$. In standard kinetic-theory language, these are also known as the Relaxation-Time Approximation (RTA) and the angular-diffusion approximation (or Fokker--Planck limit), respectively. The prefactors have been chosen so that the nonhydrodynamic gap is $\tau^{-1}$ in both models. In particular, the two models have the same ordinary diffusion coefficient and, more generally, the same $\mathfrak{D}(i\omega,0)$. Their differences at finite wavenumber therefore arise entirely from the different relaxation rates of the higher angular harmonics.

\subsection{Plotting strategy}

\begin{figure}[h!]
    \centering
\includegraphics[width=0.46\linewidth]{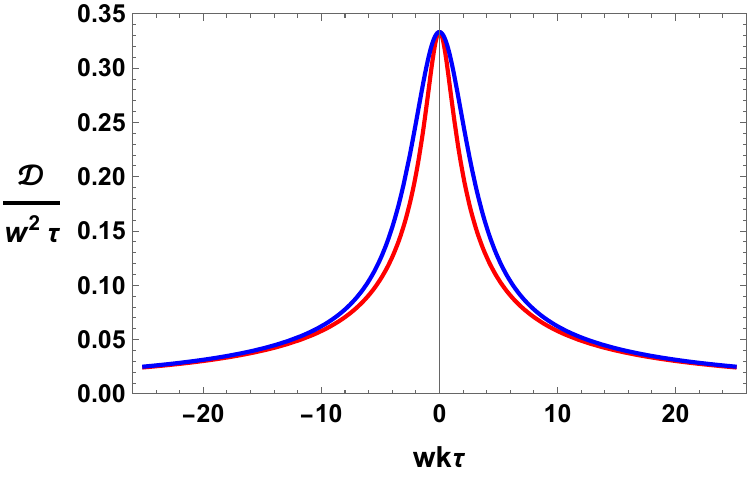}\hspace{0.6cm}\includegraphics[width=0.46\linewidth]{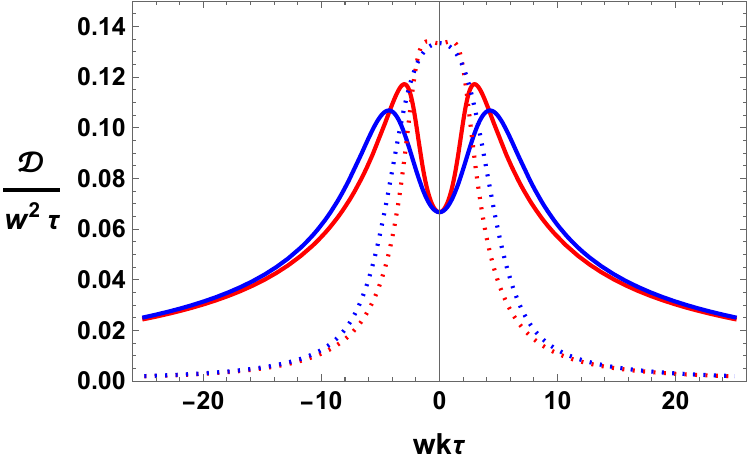}
\caption{Comparison between the resummed diffusivities for isotropic scattering (red) and small-angle scattering (blue). \textit{Left panel}: Resummed diffusivities as functions of $k\in\mathbb{R}$ at $\omega=0$, describing the flux response to a density gradient generated by a stationary source. \textit{Right panel}: Real (continuous) and imaginary (dotted) parts of the resummed diffusivities as functions of $k\in\mathbb{R}$ at $\omega=2/\tau$, describing the flux response to a density gradient generated by a source oscillating with period $\pi\tau$. The two resummed diffusivities agree at $k=0$, where only the first angular relaxation rate $\sigma_1=\tau^{-1}$ enters.}
\label{fig:Comparisonomega}
\end{figure}

The eigenvalues of $\s_N$ are known analytically in both models and are, respectively, $\sigma_\ell=\tau^{-1}$ and $\sigma_\ell=\ell(\ell+1)/(2\tau)$. To evaluate the resummed diffusivities numerically, we introduce the truncated quantities $\mathfrak{D}_\ell(i\omega,ik)$, obtained from \eqref{Diomik} by setting $a_{\ell+1}=0$. We then plot $\mathfrak{D}_\ell$ for increasing values of $\ell$ until convergence is reached. Convergence is considerably faster for small-angle scattering, owing to the progressive growth of $\sigma_\ell$. Nevertheless, we find that by $\ell\sim 1000$ both models have converged to a stable value\footnote{In the isotropic-scattering case, $\mathfrak{D}(i\omega,ik)$ can also be written in closed form, without recourse to a continued fraction. In Appendix~\ref{AAA}, we derive this expression and use it as an independent benchmark for the convergence of the truncated continued fraction.}. The resulting resummed diffusivities are shown in Fig.~\ref{fig:Comparisonomega}.

We see that the two models disagree at finite $k$, due to the additional information contained in the angular relaxation rates $\sigma_\ell$ of the higher harmonics. Remarkably, however, at very large $|k|$ they approach each other again. This is a general phenomenon: at sufficiently large real wavenumber, transport becomes dominated by free streaming, and essentially all scattering operators lead to the same asymptotic behavior. We now derive this universal large-$|k|$ limit.

\subsection{Free-streaming limit}

Consider the matrix element \eqref{matrixelment} at $\omega=0$ and $k>0$. We can write
\begin{equation}\label{Domegaiszero}
\mathfrak{D}(0,ik)=-i\dfrac{w}{k}\left(n^x Y_0^0,
\left[n^x_N-i\dfrac{\s_N}{kw}\right]^{-1}n^x Y_0^0\right) \, .
\end{equation}
For a bounded scattering operator, in the limit of very large $k$, we may formally regard $\s_N/(kw)$ as a positive infinitesimal $0^+$. Following the same procedure as in Appendix~\ref{AAA}, we then find
\begin{equation}
\left[n^x_N-i0^+\right]^{-1}n^xY_0^0
=
Y_0^0-
\dfrac{\dfrac{Y_0^0}{n^x-i0^+}}
{\left(Y_0^0,\dfrac{Y_0^0}{n^x-i0^+}\right)} \, .
\end{equation}
Substituting this expression into \eqref{Domegaiszero}, and using the Sokhotski--Plemelj formula, we obtain
\begin{equation}
\mathfrak{D}(0,ik)\longrightarrow
\dfrac{iw/k}
{\displaystyle\int_{-1}^{1}\frac{d\xi}{2(\xi-i0^+)}}
=
\dfrac{2w}{\pi k} \, .
\end{equation}
For negative $k$, the prescription is reversed and gives the same result with $k$ replaced by $|k|$. Hence,
\begin{equation}\label{freestreaminglimit}
\mathfrak{D}(0,ik)\xrightarrow{k\to\pm\infty}
\dfrac{2w}{\pi|k|} \, .
\end{equation}
For an unbounded $\s$ (as in the small-angle approximation), the above argument is not a priori guaranteed to hold. Indeed, in this case there is no scale at which free streaming dominates the collision operator uniformly: no matter how large $k$ is, there always exist sufficiently high angular harmonics whose relaxation rates exceed the free-streaming rate $kw$. Nevertheless, we show in Appendix~\ref{BBB} that, for the small-angle scattering operator, the eigenvalues do not grow rapidly enough to modify the leading large-$|k|$ behavior. As a result, $\mathfrak{D}(0,ik)$ still approaches the free-streaming result $2w/(\pi|k|)$.

\vspace{-0.3cm}
\subsection{Singularity structure}
\vspace{-0.3cm}

\begin{figure}[h!]
    \centering
\includegraphics[width=0.46\linewidth]{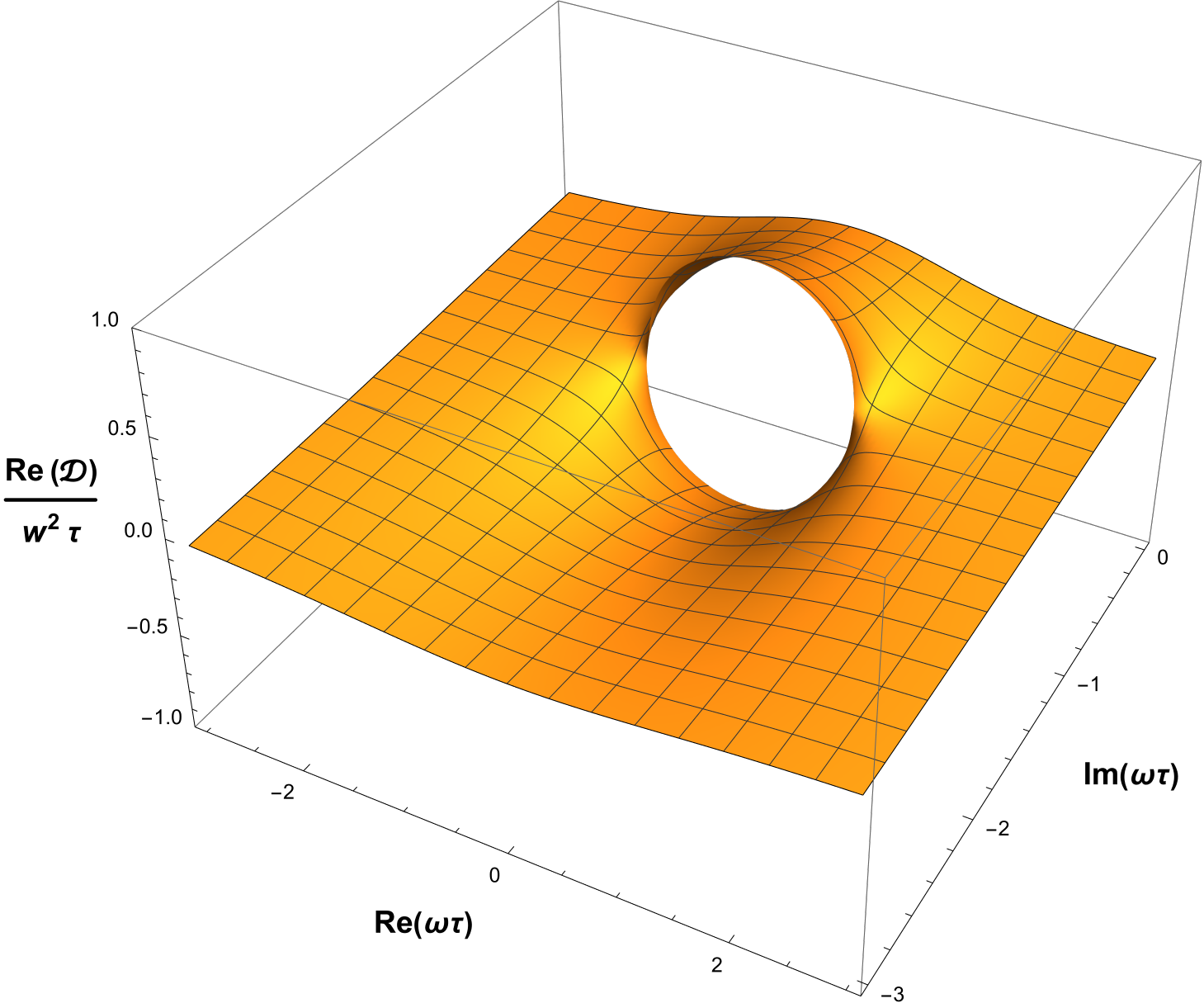}\hspace{0.6cm}\includegraphics[width=0.46\linewidth]{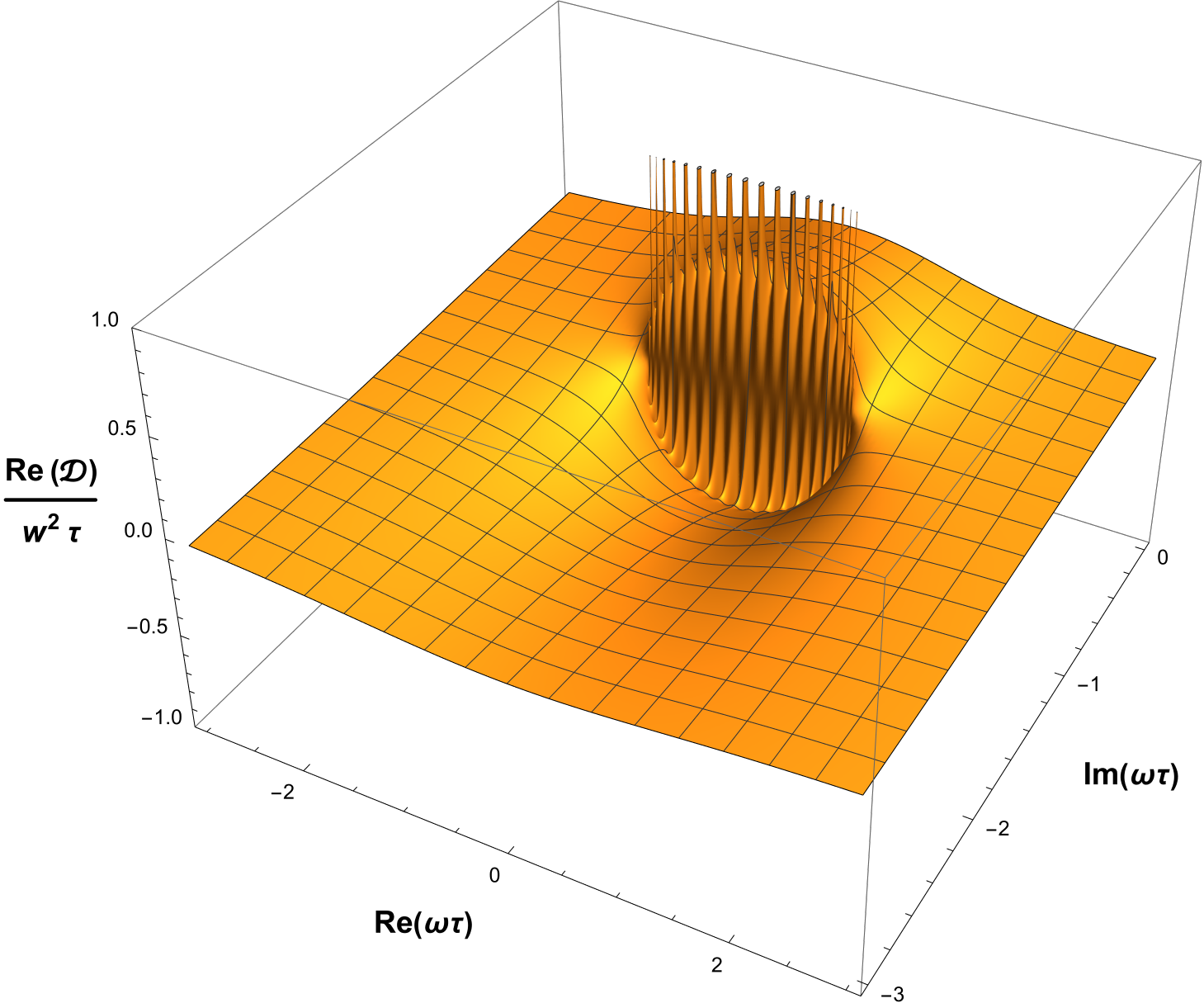}
\caption{Real part of the resummed diffusivity for isotropic scattering at $wk\tau=1$. \textit{Left panel}: Exact result, obtained from \eqref{exactD}. The branch cut along $\omega=-i/\tau+[-wk,wk]$ corresponds to the continuum of ballistic nonhydrodynamic excitations. \textit{Right panel}: Continued fraction truncated at $\ell=20$. The branch cut is replaced by a dense sequence of poles, while the truncated and exact diffusivities already agree almost perfectly away from the cut.}
\label{fig:PolesAndCut}
\end{figure}

In the companion paper, we showed that, at fixed $k$, the branch-cut singularities of $\mathfrak{D}(i\omega,ik)$ on the complex $\omega$-plane always lie on top of continuous nonhydrodynamic branches. Poles, instead, are forced to coincide with exact nonhydrodynamic modes only at $k=0$. If we truncate the continued fraction at finite $\ell$, however, we obtain a rational function, and hence a diffusivity whose singularities consist only of poles. As the truncation order is increased, the continued fraction must converge to the exact diffusivity away from the branch cut, while an increasingly dense sequence of poles needs to develop along the cut. In this way, the finite-dimensional truncations reconstruct the continuous nonhydrodynamic spectrum of the full theory. We observe this for isotropic scattering in Fig.~\ref{fig:PolesAndCut}. 

\begin{figure}[h!]
    \centering
\includegraphics[width=0.46\linewidth]{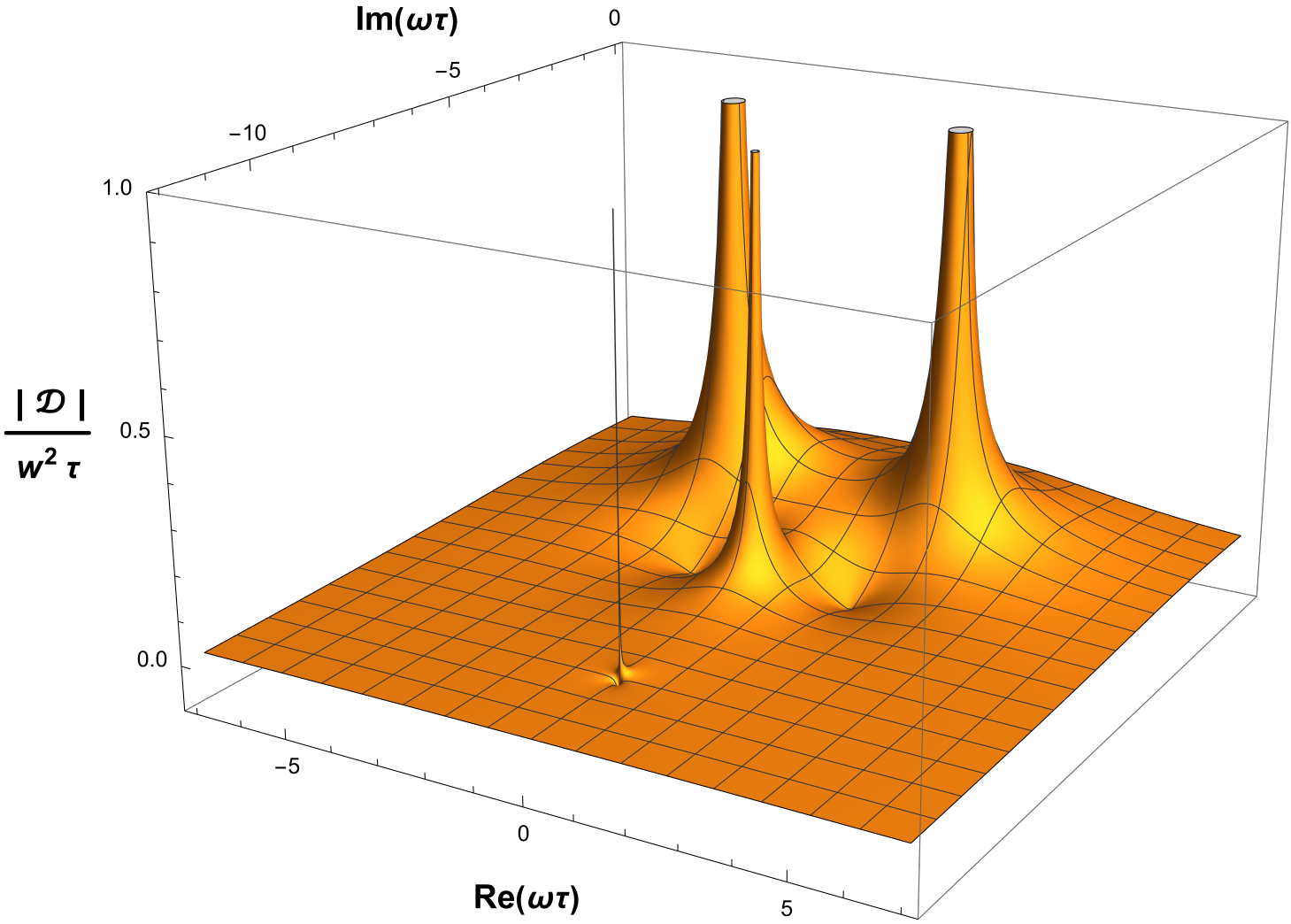}\hspace{0.6cm}\includegraphics[width=0.46\linewidth]{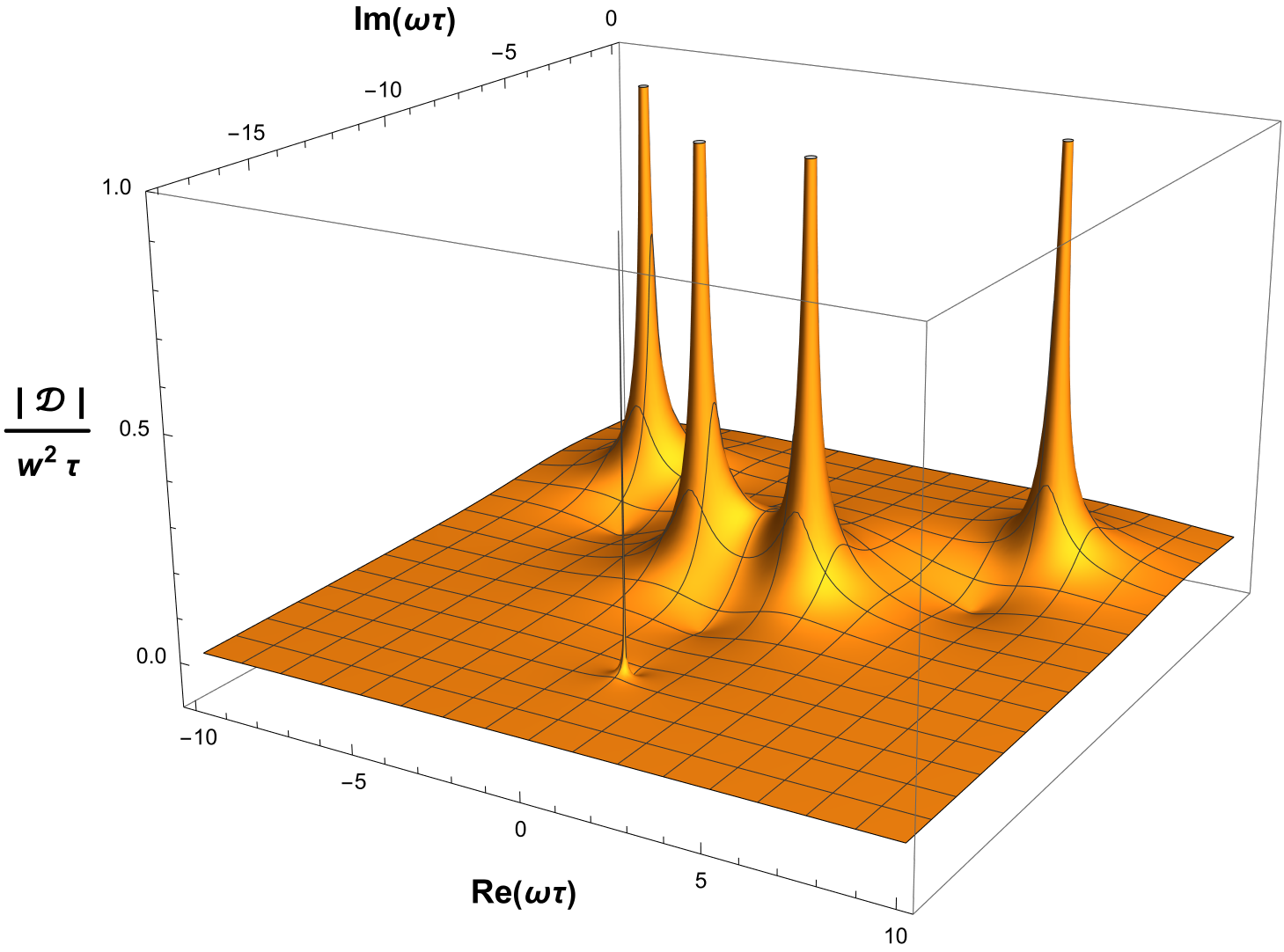}
\caption{Absolute value of the resummed diffusivity for small-angle scattering at $wk\tau=5$ (left panel) and $10$ (right panel). Both plots are evaluated by truncating the continued fraction at $\ell=20$. As $|k|$ increases, the poles collide pairwise (the first with the second, the third with the fourth, and so on) and subsequently move off the imaginary axis as conjugate pairs, signaling the transition from purely damped to propagating excitations. Since successive poles are increasingly far apart, these collisions occur at progressively larger values of $|k|$.}
\label{fig:PolesDA}
\end{figure}

The singularity structure in the small-angle-scattering limit is qualitatively different. In fact, a continuous ballistic branch can exist only if, at $k=0$, the spectrum possesses an accumulation point or an eigenvalue of infinite multiplicity \cite[Theorem 5]{GavassinoLorentzian:2026seq}. For small-angle scattering, all eigenvalues $\sigma_\ell$ have finite multiplicity and become increasingly sparse. Hence, no branch cut can arise. Physically, this means that angular diffusion destroys the continuous ballistic branch. This is reflected in the fact that $\mathfrak{D}$ retains well-separated poles also in the large-$\ell$ limit; see Fig.~\ref{fig:PolesDA}.

We remark that, although the poles of $\mathfrak{D}$ do not in general coincide with the exact nonhydrodynamic modes of the theory, they still admit a closely related interpretation. Indeed, their relaxation rates $i\omega$ belong to $\operatorname{Spectrum}(\s_N{+}ikw n_N^x)$, whereas the actual nonhydrodynamic rates belong to $\operatorname{Spectrum}(\s{+}ikw n^x)$. The difference is therefore that the former describe the spectrum obtained after the conserved density has been removed as a degree of freedom, leaving only the purely nonhydrodynamic sector.

\subsection{Recovering the modes from the resummed diffusivity}

Even though the poles of the resummed diffusivity are not, in general, genuine excitation modes of the theory, the true modes can still be reconstructed directly from it. In fact, suppose that $\omega(k)$ is an isolated excitation mode of the system and that it does not coincide with a pole of $\mathfrak{D}$. We can then write the conservation law as
$0=-i\omega\rho+ikJ=-i\omega\rho-(ik)^2\mathfrak{D}(i\omega,ik)\rho$,
where the second equality is allowed because $\mathfrak{D}$ is well defined there. If the mode carries a nonvanishing particle-density perturbation, $\rho\neq0$, its dispersion relation must therefore satisfy
\begin{equation}\label{dispersionD}
k^2\mathfrak{D}(i\omega,ik)-i\omega=0 \, .
\end{equation}
Thus, the resummed diffusivity allows us to recover not only the hydrodynamic mode, but also all isolated excitation modes that couple to the particle density and do not coincide with its singularities.\footnote{Note that non-hydrodynamic modes associated with spherical harmonics $Y_\ell^m$ with $m\neq0$ carry zero particle density and therefore cannot be reconstructed from $\mathfrak{D}$.}

\begin{figure}[h!]
    \centering
\includegraphics[width=0.4\linewidth]{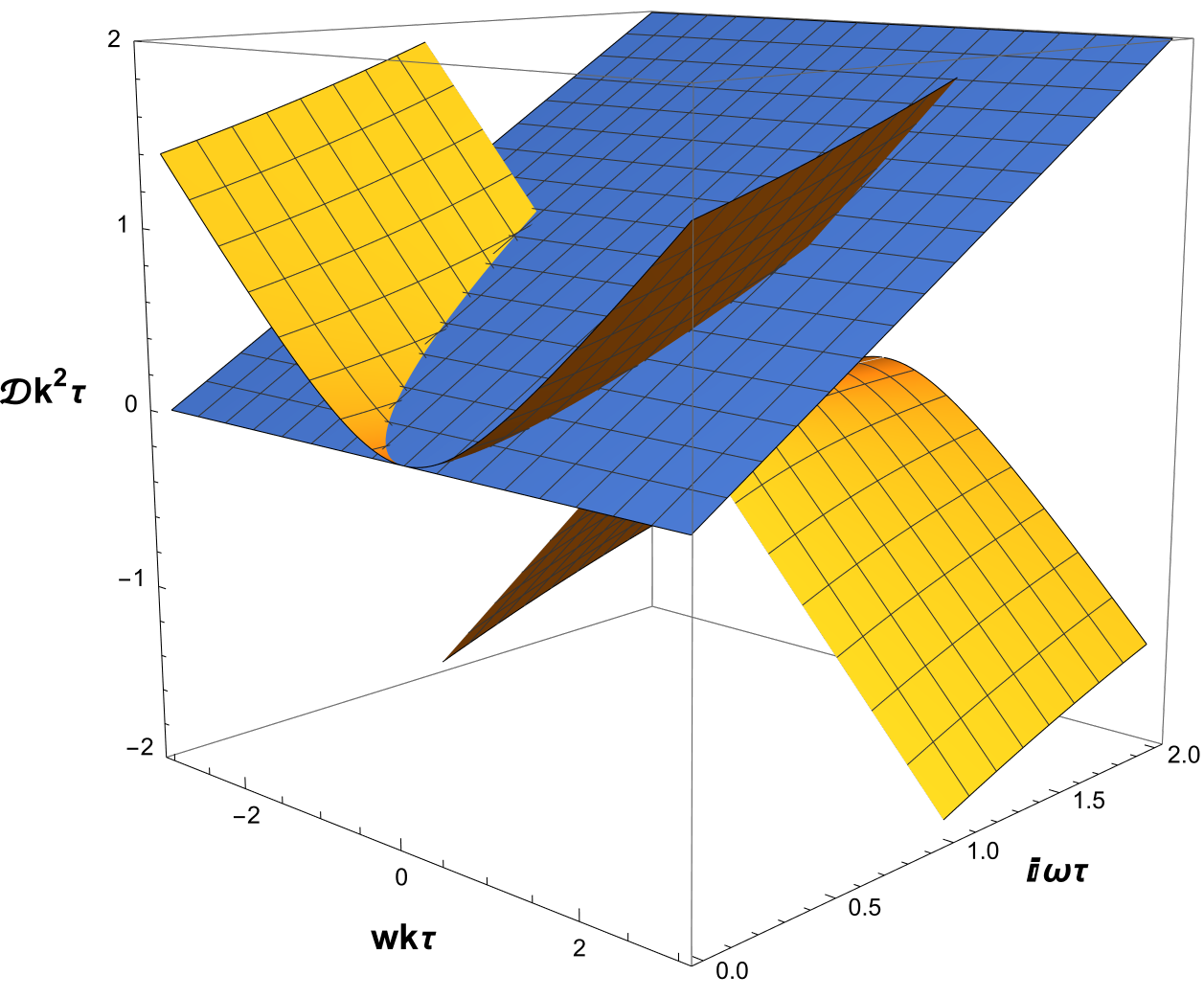}\hspace{0.6cm}\includegraphics[width=0.47\linewidth]{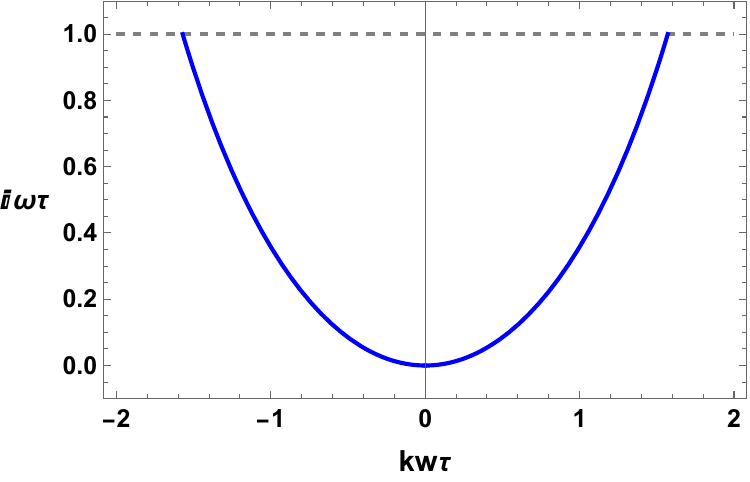}
\includegraphics[width=0.4\linewidth]{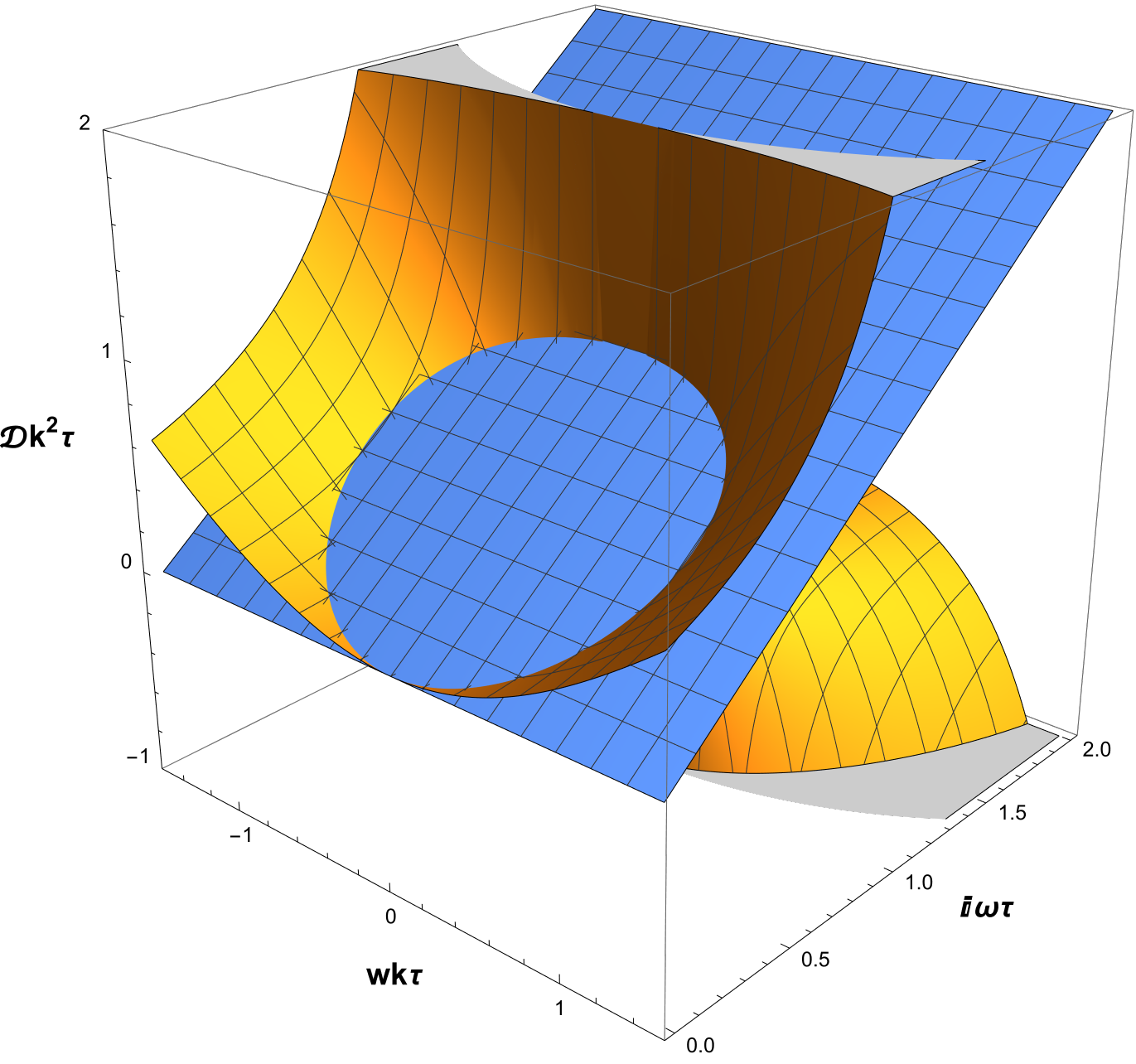}\hspace{0.6cm}\includegraphics[width=0.47\linewidth]{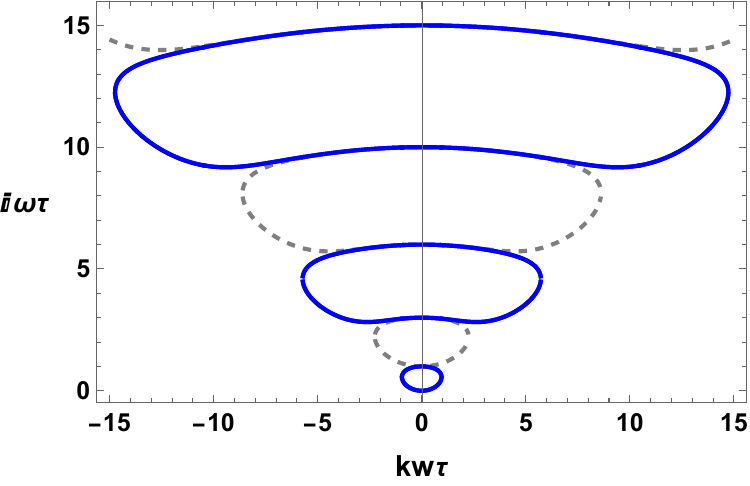}
\caption{Graphical reconstruction of the dispersion relations for isotropic scattering (upper panels) and small-angle scattering (lower panels).
\textit{Upper-left panel}: The function $\mathfrak{D}k^2\tau$ (orange) grows as $\sim k^2$ for small $k$. Hence, its intersection with the plane $i\omega\tau$ (blue) is approximately parabolic. As $i\omega\tau$ approaches $1$, however, $\mathfrak{D}$ encounters the branch cut associated with the continuum of ballistic excitations. Across the cut, $\mathfrak{D}k^2\tau$ changes sign, and no further intersection exists on the real $\{i\omega,k\}$ plane. The hydrodynamic mode therefore terminates at the cut, without giving rise to an isolated non-hydrodynamic mode. The resulting dispersion relation is shown in the upper-right panel, where the gray dashed line marks the branch cut location.
\textit{Lower-left panel}: At small $i\omega\tau$, the behavior of $\mathfrak{D}k^2\tau$ is analogous to the isotropic-scattering case, producing the parabolic hydrodynamic dispersion relation. In this case, however, $\mathfrak{D}$ has no branch cuts and instead diverges at a discrete set of poles (gray dashed curves in the lower-right panel). Consequently, an infinite tower of non-hydrodynamic modes emerges from $i\omega\tau=\sigma_\ell\tau=\ell(\ell+1)/2$ at $k=0$. These modes cannot remain non-propagating at arbitrarily large $|k|$, since $\mathfrak{D}k^2\tau\sim |k|$ in the free-streaming limit \eqref{freestreaminglimit}, whereas $i\omega\tau$ remains $k$-independent on any horizontal line. They must therefore collide pairwise and subsequently leave the real $\{i\omega,k\}$ plane. The collision between the hydrodynamic mode and the first non-hydrodynamic mode is analogous to that occurring in Cattaneo's theory.}
\label{fig:DispesionIntersection}
\end{figure}

For non-propagating Fourier modes (for which $k\in\mathbb{R}$ and $i\omega\in\mathbb{R}$), the left-hand side of \eqref{dispersionD} is a real-valued function on the $\{i\omega,k\}$ plane. Hence, away from degenerate roots, its zero set generically consists of curves, which describe the dispersion relations of the non-propagating modes. The implicit function theorem then allows us to constrain how these curves can evolve and terminate. Moreover, even when $\mathfrak{D}$ has a cumbersome analytic form, the structure of these solutions can be inferred graphically by comparing $k^2\mathfrak{D}(i\omega,ik)$ with $i\omega$, as shown in Fig.~\ref{fig:DispesionIntersection}.

Note that, for isotropic scattering, the dispersion relation can be obtained analytically from \eqref{exactD}, and coincides with the charge-density mode of massless RTA kinetic theory \cite{RomatschkeCutsandPoles:2015gic}:
\begin{equation}
i\omega\tau
=1-\dfrac{kw\tau}{\tan(kw\tau)} \qquad \qquad kw\tau \in \left(-\dfrac{\pi}{2}, \dfrac{\pi}{2}\right) \, .
\end{equation}

\vspace{-0.3cm}
\section{Realistic scattering integrals}
\vspace{-0.3cm}

The two scattering models considered so far represent opposite limits of a much broader class of realistic collision operators \cite{case1967linear,pomraning1973equations,Levermore1984}, which take the general form
\begin{equation}\label{realisticintegral}
(\s I)(\n)=\int d^2\Omega'\,F(\n{\cdot}\n')
\left[I(\n)-I(\n')\right] \, ,
\end{equation}
where $F$ is the scattering redistribution function, proportional to the differential scattering cross section. In this section, we consider a family of functions $F_b$, with $b\in[0,+\infty)$, whose associated dynamics interpolates between isotropic scattering at $b=0$ and small-angle scattering as $b\to\infty$, while keeping the non-hydrodynamic gap fixed. This will allow us to understand how $\mathfrak{D}(i\omega,ik)$ behaves in the intermediate regime, where a ballistic branch cut coexists with a set of isolated poles.

\subsection{Interpolating model}

We take
\begin{equation}\label{Fbexponential}
F_b(\n{\cdot}\n')
=
\dfrac{e^{b\n\cdot\n'}}
{4\pi\tau\left[
(1{+}\frac{1}{b})\frac{\sinh(b)}{b}
-\frac{\cosh(b)}{b}
\right]} \, .
\end{equation}
For $b\to0$, we have $F_b\to(4\pi\tau)^{-1}$, and we recover isotropic scattering. By contrast, as $b\to+\infty$, the scattering kernel becomes increasingly peaked around $\n=\n'$, corresponding to the small-angle limit.

\begin{figure}[h!]
    \centering
\includegraphics[width=0.46\linewidth]{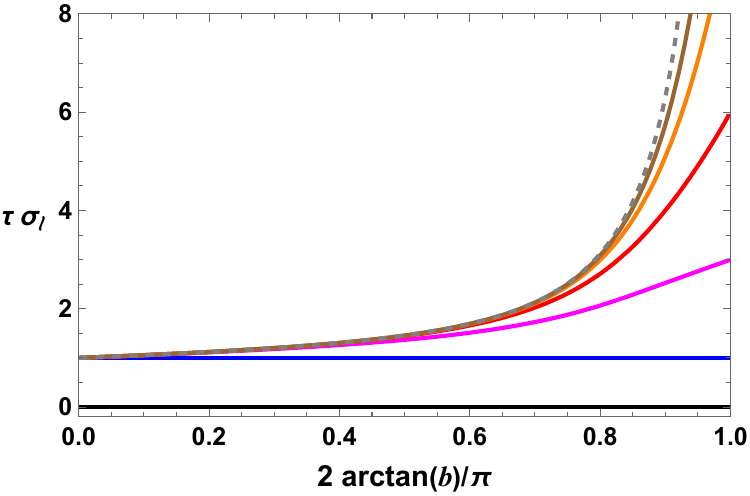}
\caption{Eigenvalues of the interpolating collision operator as $b$ is varied. The curves correspond to $\ell=0$ (black), $1$ (blue), $2$ (magenta), $3$ (red), $4$ (orange), and $5$ (brown), while the dashed curve denotes the accumulation point $\sigma_\infty$. Infinitely many additional eigenvalues lie between the brown and dashed curves and accumulate at $\sigma_\infty$ as $\ell\to\infty$. The normalization of the scattering integral has been chosen so that $\sigma_1=\tau^{-1}$ for all $b$. Consequently, all models have the same resummed diffusivity at zero wavenumber, $\mathfrak{D}(i\omega,0)=\frac{w^2\tau}{3(1-i\omega\tau)}$, and differ only at finite $k$.}
\label{fig:Varyingb}
\end{figure}

Integrating \eqref{realisticintegral} over all directions, we immediately obtain $\sigma_0=0$ for any $F$, in agreement with particle-number conservation. For the higher harmonics, rotational invariance gives
\begin{equation}
\sigma_\ell
=
\left.
\dfrac{(\s Y_\ell^0)(\n)}{Y_\ell^0(\n)}
\right|_{\n=\text{north pole}}
=
2\pi\int_{-1}^1d\xi\,F_b(\xi)
\left[1-\mathcal{P}_\ell(\xi)\right]
=
\dfrac{
\frac{\sinh(b)}{b}-i_\ell(b)
}{
(1{+}\frac{1}{b})\frac{\sinh(b)}{b}
-\frac{\cosh(b)}{b}
}\,\tau^{-1} \, ,
\end{equation}
where $\mathcal{P}_\ell$ are the Legendre polynomials and $i_\ell$ are the modified spherical Bessel functions of the first kind \cite[Eq.~(C.39)]{Theeten2009}. The behavior of the eigenvalues is shown in Fig.~\ref{fig:Varyingb}.

For any fixed $b<\infty$, $i_\ell(b)\to0$ as $\ell\to\infty$. Hence, the infinite sequence of eigenvalues accumulates at
\begin{equation}
\sigma_\infty
=
\dfrac{\tau^{-1}}
{1+\frac{1}{b}-\coth(b)} \, .
\end{equation}
This accumulation point is not itself an eigenvalue, but belongs to the essential spectrum. Moreover, $\s-\sigma_\infty$ is compact. It follows that, at finite wavenumber, the corresponding essential spectrum is broadened by free streaming into a ballistic branch extending over $\omega=-i\sigma_\infty+[-wk,wk]$ \cite{GavassinoLorentzian:2026seq}. Thus, a ballistic branch cut is present for every finite $b$ and $k\neq0$. As $b\to\infty$, however, $\sigma_\infty\to\infty$, so that the cut is pushed arbitrarily far down the complex-frequency plane. Only in the strict $b=\infty$ limit does it disappear to infinity, leaving behind the discrete spectral structure of the small-angle scattering operator.

Indeed, the small-angle eigenvalues are recovered directly in this limit. For fixed $\ell$ and large $b$, the scattering kernel is localized near $\xi=1$, where $\mathcal{P}_\ell(\xi)
=
1+\frac{\ell(\ell+1)}{2}(\xi-1)
+\mathcal{O}\big((1-\xi)^2\big)$.
Therefore,
\begin{equation}
\sigma_\ell
=
\dfrac{\tau^{-1}}{2}
\int_{-1}^1d\xi\,
\dfrac{e^{b\xi}\left[1-\mathcal{P}_\ell(\xi)\right]}
{(1{+}\frac{1}{b})\frac{\sinh(b)}{b}
-\frac{\cosh(b)}{b}}
\simeq
\dfrac{\ell(\ell+1)}{2\tau}
\int_{-1}^1d\xi\,
\dfrac{b^2e^{b\xi}(1-\xi)}{e^b}
\longrightarrow
\dfrac{\ell(\ell+1)}{2\tau} \, ,
\end{equation}
as required.

\subsection{Singularity structure}
\vspace{-0.3cm}

\begin{figure}[h!]
    \centering
\includegraphics[width=0.37\linewidth]{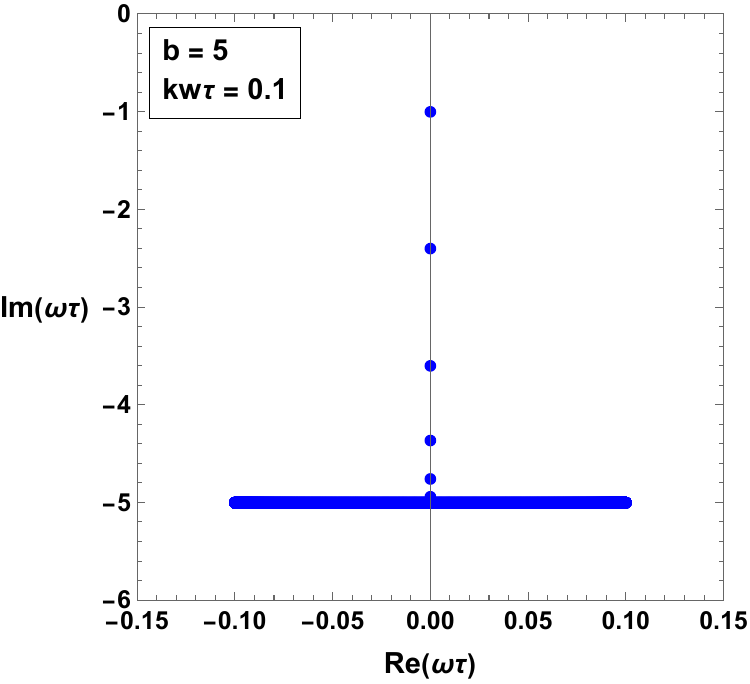}\hspace{0.6cm}\includegraphics[width=0.37\linewidth]{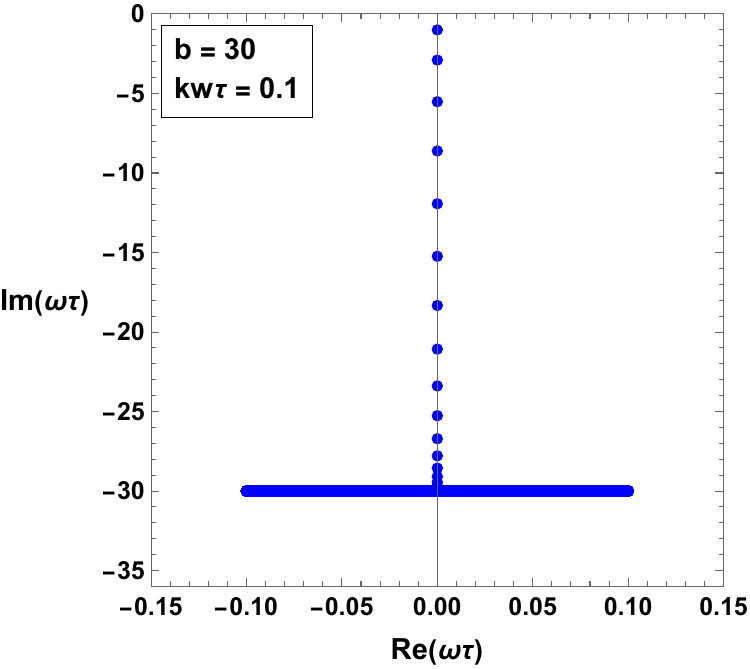}
\includegraphics[width=0.38\linewidth]{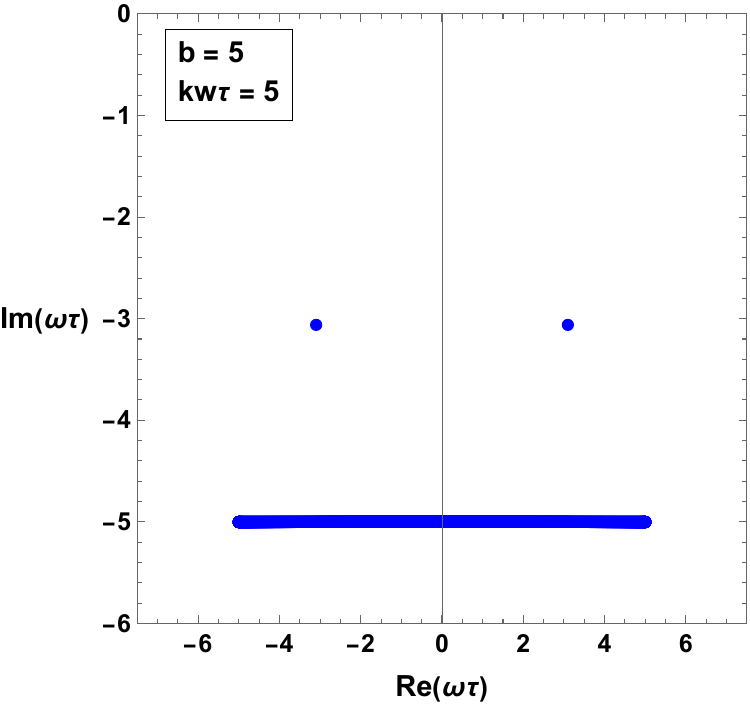}\hspace{0.6cm}\includegraphics[width=0.38\linewidth]{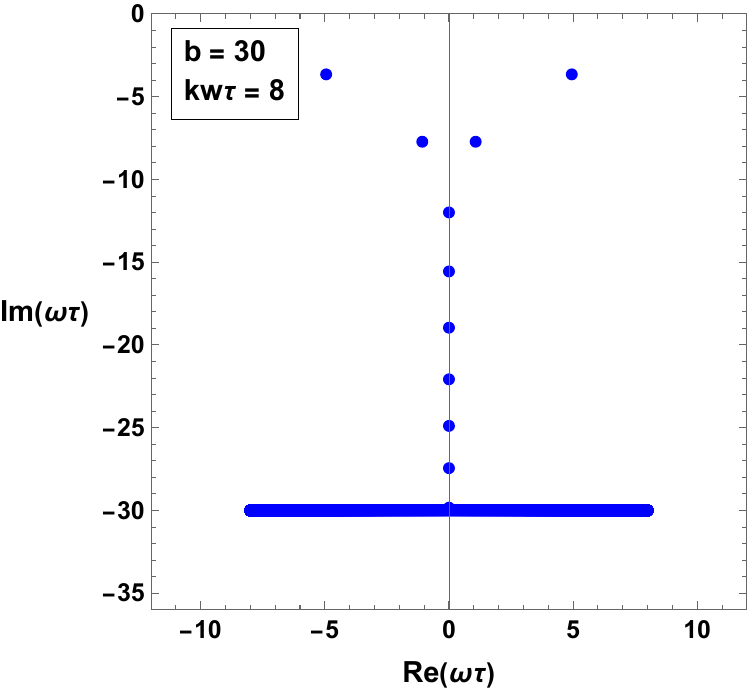}
\includegraphics[width=0.37\linewidth]{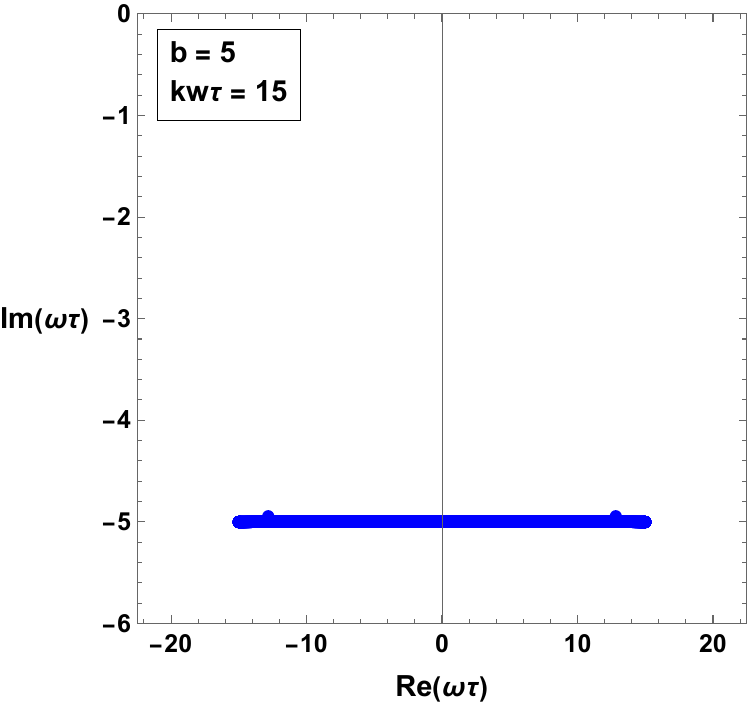}\hspace{0.6cm}\includegraphics[width=0.38\linewidth]{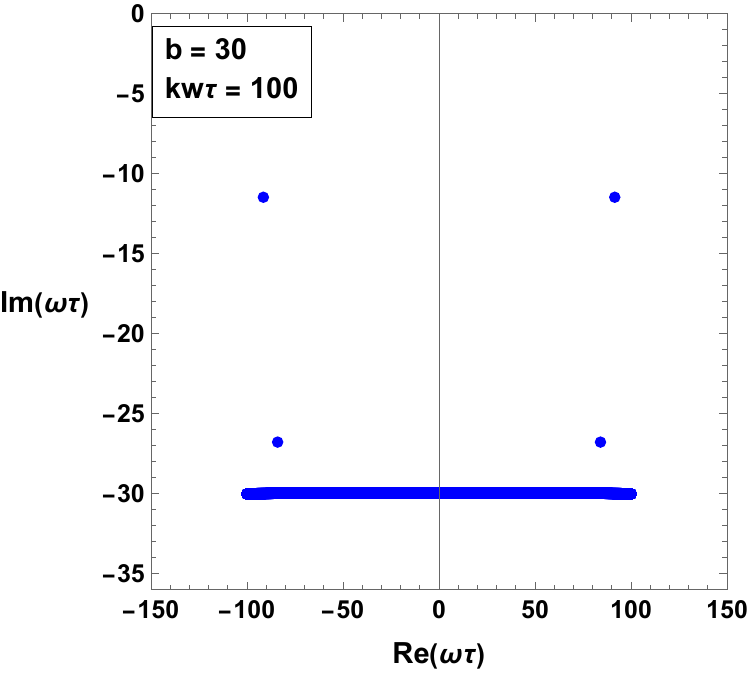}
\caption{Singularities of $\mathfrak{D}(i\omega,ik)$, truncated at the $3000$th spherical harmonic, for $b\,{=}\,5$ (left column) and $b\,{=}\,30$ (right column), at different values of $k$. The large number of closely spaced poles reconstructs the expected ballistic branch cut, as in Fig.~\ref{fig:PolesAndCut}, while several additional isolated poles are visible away from it. Most of these poles (infinitely many, in the $\ell\to\infty$ limit) merge into the cut soon after a finite $k$ is turned on. Those closest to the origin, instead, can first collide pairwise and move off the imaginary axis, before eventually merging into the cut at sufficiently large $|k|$.}
\label{fig:RealisticIntermedio}
\end{figure}

The poles of the truncated diffusivity $\mathfrak{D}_\ell(i\omega,ik)$ coincide, up to pole skipping (which is non-generic at finite $k$), with the eigenvalues of $\s_N+ikw n_N^x$ restricted to $\operatorname{span}\{Y_1^0,\ldots,Y_\ell^0\}$. Numerically, it is more convenient to compute these eigenvalues directly, rather than to search for the zeros of $1/\mathfrak{D}_\ell$. The result for $\ell=3000$ is shown in Fig.~\ref{fig:RealisticIntermedio}.

\subsection{Exact solution for first-order anisotropic scattering}
\vspace{-0.3cm}

The excitation modes of the kinetic theory exhibit a pattern qualitatively similar to that observed for the poles of $\mathfrak{D}$ in Fig.~\ref{fig:RealisticIntermedio}. In particular, as $k$ increases, the hydrodynamic mode can undergo two different fates: it can merge directly into the ballistic cut, as in the isotropic-scattering case, or it can first collide with a non-hydrodynamic mode and become propagating, as in the small-angle scattering limit, before eventually being absorbed into the cut. It is natural to ask how the transition between these two behaviors occurs. Rather than analyzing the model \eqref{Fbexponential} numerically, it is more instructive to consider a simpler model, commonly used in radiative-transfer applications, for which the dispersion relation can be obtained analytically as parametric curves.

\begin{figure}[h!]
    \centering
\includegraphics[width=0.45\linewidth]{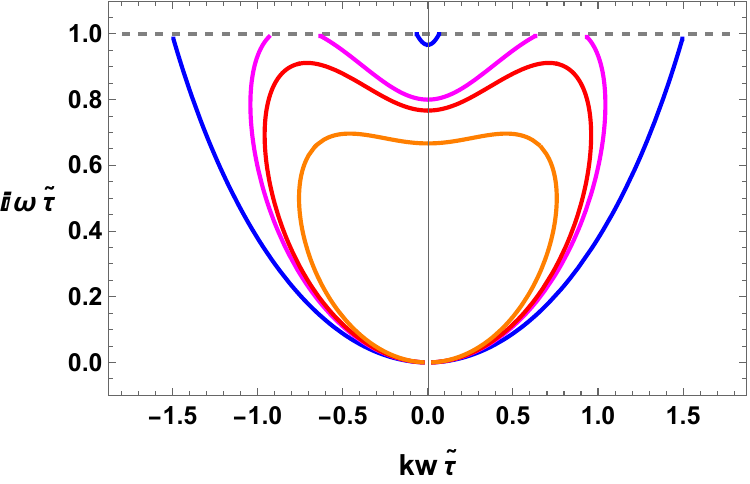}
\caption{Dispersion relations of the non-propagating modes in the first-order anisotropic scattering approximation, for $b=0.1$ (blue), $0.6$ (magenta), $0.7$ (red), and $1$ (orange). The dashed line denotes the branch cut. Three distinct regimes can be identified. For $b\lesssim0.44$, both the hydrodynamic and the non-hydrodynamic mode merge directly into the cut. For $0.44\lesssim b\lesssim0.62$, the non-hydrodynamic mode first merges into the cut, but subsequently reemerges from the continuum and collides with the hydrodynamic mode, producing a propagating pair (not shown), which is eventually reabsorbed into the cut. Finally, for $b\gtrsim0.62$, the two modes collide before reaching the cut, producing a propagating pair that is eventually absorbed into it.}
\label{fig:Firstorderanis}
\end{figure}

We work in the so-called ``first-order anisotropic approximation'', whose redistribution function is
\begin{equation}
F_b(\n{\cdot}\n')
=
\dfrac{1+b\,\n{\cdot}\n'}
{4\pi\tau(1{-}\frac{b}{3})}
\qquad\qquad
(0\leq b\leq1) \, .
\end{equation}
This agrees with \eqref{Fbexponential} to first order in $b$, while giving rise to a much simpler spectrum of the scattering operator:
\vspace{-0.1cm}
\begin{equation}
\sigma_\ell=\dfrac{1}{\tau}
\begin{cases}
0\, , & \ell=0\, , \\
1\, , & \ell=1\, , \\
(1{-}\frac{b}{3})^{-1}\, , & \ell\geq2 \, .
\end{cases}
\end{equation}
The fact that the eigenvalues are constant for all $\ell\geq2$ allows us to evaluate the resummed diffusivity exactly (see Appendix~\ref{CCC}), without recourse to a continued fraction:
\vspace{-0.1cm}
\begin{equation} \mathfrak{D}(i\omega,ik)=\dfrac{\Tilde{\tau} w^2}{\dfrac{(ikw\Tilde{\tau})^2}{1{-}i\omega\Tilde{\tau}} \left[ 1-\dfrac{\frac{ikw\Tilde{\tau}}{1-i\omega\Tilde{\tau}}} {\operatorname{arctanh}\left(\frac{ikw\Tilde{\tau}}{1-i\omega\Tilde{\tau}}\right)} \right]^{-1}-b}\, , \qquad\qquad \Tilde{\tau}=\left(1-\frac{b}{3}\right)\tau \, .
\end{equation}
Substituting this expression into the conservation law $-i\omega-(ik)^2\mathfrak{D}=0$, and introducing
$r=kw\Tilde{\tau}/(1-i\omega\Tilde{\tau})\in\mathbb{R}$, we obtain two parametric curves describing the non-propagating portions of the hydrodynamic mode and of the isolated non-hydrodynamic mode that originates from the $\ell=1$ eigenvalue of $\s$. Explicitly,
\vspace{-0.1cm}
\begin{equation} \begin{cases} (kw\Tilde{\tau})_\pm=\dfrac{b r-b\, \text{arctan}(r)+r^2 \text{arctan}(r)\pm\sqrt{4 b r^3 \left[\text{arctan}(r)-r\right]+\left[\left(r^2{-}b\right) \text{arctan}(r)+b r\right]^2}}{2 r^2} \, , \\ (i\omega\Tilde{\tau})_\pm=1-\dfrac{(kw\Tilde{\tau})_\pm}{r} \, . \\ \end{cases} \end{equation}
The resulting behavior of the two modes is shown in Fig.~\ref{fig:Firstorderanis}.

\section{Frequency-dependent form factors}

Throughout this article, we have discussed extensively the properties of $\mathfrak{D}(i\omega,ik)$. So far, however, we have used it concretely only to compute dispersion relations. The main advantage of knowing $\mathfrak{D}(i\omega,ik)$ itself, rather than just the dispersion relations, is that the resummed diffusivity also determines the response of the medium in the presence of an external source, which is the context in which it is most naturally defined (see also the companion paper). We now discuss a concrete application in this direction.

\vspace{-0.3cm}
\subsection{Formal definition of the form factor}
\vspace{-0.3cm}

Consider again the Boltzmann equation \eqref{boltzmann}, this time allowing for gradients in all spatial directions, and set $Q=e^{-i\omega t}\delta^3(\mathbf{x})$, corresponding to a source that is periodic in time and localized in space (for example, an emission hot spot that is periodically heated and cooled). Then, imposing that the solution is also periodic in time, we have
\begin{equation}\label{boltzmannAGAIN}
(\s-i\omega+wn^j\partial_j)(\rho Y_0^0+I_N)
=
e^{-i\omega t}\delta^3(\mathbf{x})Y_0^0 \, .
\end{equation}
Following the same steps as in Sec.~\ref{resummeddiffusivitygneeraltheory}, we obtain
\begin{equation}
\rho(t,\mathbf{x})
=
e^{-i\omega t}
\int \dfrac{d^3k}{(2\pi)^3}
\dfrac{e^{i\mathbf{k}\cdot\mathbf{x}}}
{(w\PN\mathbf{k}{\cdot}\mathbf{n}Y_0^0,
[\s_N+wi\mathbf{k}{\cdot}\mathbf{n}_N-i\omega]^{-1}
w\PN\mathbf{k}{\cdot}\mathbf{n}Y_0^0)-i\omega}
\, .
\end{equation}
Due to rotational invariance, the matrix element in the denominator equals $k^2\mathfrak{D}(i\omega,ik)$, with $k=|\mathbf{k}|$. Going to spherical coordinates and defining $r=|\mathbf{x}|$, the above integral simplifies to
\begin{equation}
\rho(t,r)
=
\dfrac{e^{-i\omega t}}{4\pi r}
\int_{-\infty}^\infty \dfrac{dk}{i\pi}
\dfrac{k e^{ikr}}
{k^2\mathfrak{D}(i\omega,ik)-i\omega}
\, .
\end{equation}
In the Fick limit, where $\mathfrak{D}$ is constant, the integral evaluates to $\frac{1}{\mathfrak{D}}\exp[-\sqrt{-i\omega r^2/\mathfrak{D}}]$. It is therefore natural to factor out the Fick prediction and write
\begin{equation}
\rho(t,r)
=
\dfrac{e^{-i\omega t-\sqrt{\frac{-i\omega r^2}{\mathfrak{D}(0,0)}}}}
{4\pi\mathfrak{D}(0,0)r}
f(\omega,r) \, ,
\end{equation}
where the form factor $f(\omega,r)$ can be expressed directly as the ratio
\begin{equation}\label{fomegatdefined}
f(\omega,r)
=
\dfrac{\displaystyle\int_{-\infty}^\infty
\dfrac{k e^{ikr} dk}
{k^2\mathfrak{D}(i\omega,ik)-i\omega}}
{\displaystyle\int_{-\infty}^\infty
\dfrac{k e^{ikr} dk}
{k^2\mathfrak{D}(0,0)-i\omega}}
\, .
\end{equation}
Thus, $f-1$ quantifies the deviation of the response from the prediction of Fick's law. Since $f$ is entirely determined by the resummed diffusivity, it provides an observable probe of its frequency- and wavenumber-dependence. In particular, the dependence on $\omega$ probes temporal nonlocality, whereas the dependence on $r$ probes the finite-wavenumber structure of the constitutive relation.

\vspace{-0.3cm}
\subsection{Numerical evaluation}
\vspace{-0.3cm}

\begin{figure}[h!]
    \centering
\includegraphics[width=0.45\linewidth]{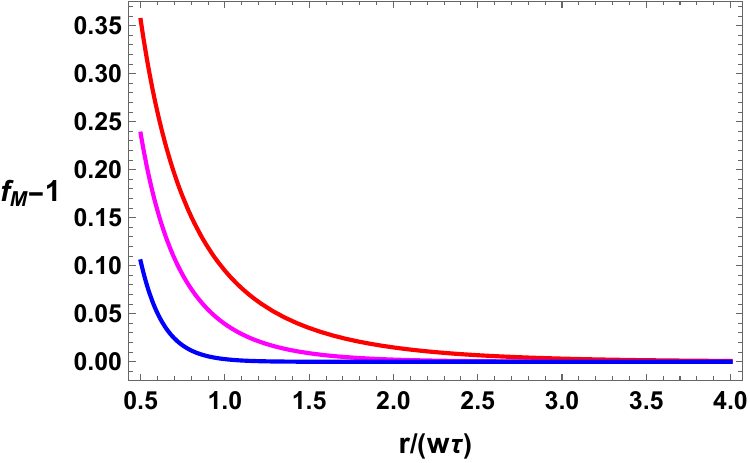}\hspace{0.6cm}\includegraphics[width=0.45\linewidth]{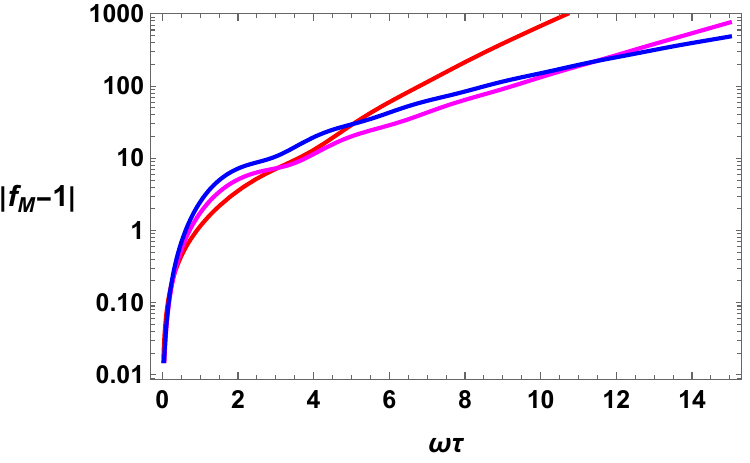}
\caption{Mollified form factor for isotropic scattering (red), the realistic scattering kernel \eqref{Fbexponential} with $b=2$ (magenta), and small-angle scattering (blue). \textit{Left panel}: Spatial dependence at $\omega=0$, where $f_M$ is real. The deviation $f_M-1$ decreases rapidly with the distance from the source, with a decay rate that depends strongly on the scattering kernel. \textit{Right panel}: Frequency dependence at $r=3w\tau$. We plot $|f_M-1|$, since the form factor is oscillatory at large frequency: the kinetic and Fick responses acquire different phases and periodically move in and out of phase. The deviation from Fick's law grows strongly with frequency, reflecting the increasing importance of the ballistic response relative to the exponentially suppressed diffusive prediction. At sufficiently large $\omega\tau$, the result eventually becomes sensitive to the mollification scale. Indeed, the characteristic wavelengths probed by the response become comparable to or smaller than $r_0$, so that the measured density effectively averages over different points within the spatially extended source.}
\label{fig:FormFactorMollified}
\end{figure}

Due to the free-streaming decay law \eqref{freestreaminglimit}, the integrand in \eqref{fomegatdefined} does not decay to zero at large $|\xi|$. Instead, it remains oscillatory, preventing convergence in the ordinary sense. Mathematically, this is not an issue, since the original Fourier integral is to be understood in a distributional sense. Nevertheless, for numerical purposes, it is convenient to work with an integrand that decays at infinity. We therefore mollify the source in \eqref{boltzmannAGAIN}, replacing the point-like source by a normalized Gaussian of width $r_0$. In Fourier space, this amounts to multiplying the source by $e^{-r_0^2 k^2}$, so that the form factor becomes
\begin{equation}\label{fomegatmollidied}
f_M(\omega,r)
=
\dfrac{\displaystyle\int_{-\infty}^\infty
\dfrac{k e^{ikr-r_0^2 k^2} dk}
{k^2\mathfrak{D}(i\omega,ik)-i\omega}}
{\displaystyle\int_{-\infty}^\infty
\dfrac{k e^{ikr-r_0^2k^2} dk}
{k^2\mathfrak{D}(0,0)-i\omega}}
\end{equation}
As our mollification scale, we take $r_0/(w\tau)\sim 10^{-1}$, so that the source is effectively point-like on scales $r\gtrsim w\tau/10$ and $\omega\tau \lesssim 10$. The result is shown in Fig. \ref{fig:FormFactorMollified}.

\subsection{Large-$r$ asymptotics}

By studying the behavior of $f$ at large $r$, one can infer rigorous information about the analytic structure of $\mathfrak{D}(i\omega,ik)$, and therefore about the scattering integral, similarly to how studying the late-time behavior of Bjorken solutions gives access to the non-hydrodynamic modes \cite{FlorkowskiReview2018}. 

We shall focus here on the large-$r$ limit for $\omega=0$. In this case, we have
\begin{equation}\label{fomegatmollidied2}
f_M(0,r)-1
=
\dfrac{\displaystyle\int_{-\infty}^\infty
\dfrac{dk}
{k} e^{ikr-r_0^2 k^2} \left[\dfrac{1}{\mathfrak{D}(0,ik)}-\dfrac{1}{\mathfrak{D}(0,0)} \right]}
{\displaystyle\int_{-\infty}^\infty
\dfrac{e^{ikr-r_0^2k^2} dk}
{k\mathfrak{D}(0,0)}} \, .
\end{equation}
Now, along lines $\mathrm{Im}(k)=\eta$, the mollification factor satisfies $|e^{-r_0^2k^2}|=e^{-r_0^2(\mathrm{Re}k)^2+r_0^2\eta^2}$, and therefore remains Gaussian-damped as $|\mathrm{Re}k|\to\infty$, while the square bracket grows at most like $|\mathrm{Re}k|$, due to \eqref{freestreaminglimit}. Hence, we can translate the integration contour from the real axis to $\mathrm{Im}(k)=\eta$, as long as no singularity is crossed. Since $e^{ikr}=e^{i(\mathrm{Re}k)r-\eta r}$ along the translated contour, the Fourier integral in the numerator decays at least like $e^{-(\eta_{\text{max}}-0^+)r}$, where $\eta_{\text{max}}$ corresponds to the boundary of the strip within which the integrand remains analytic. Considering that the integrand is regular at $k=0$ (the factor $1/k$ being canceled by the difference of $1/\mathfrak{D}$), we conclude that $\eta_\text{max}>0$ is determined by the first singularity of $1/\mathfrak{D}(0,ik)$ encountered upon moving away from the real $k$ axis. Such a singularity may originate from a branch cut of the resummed diffusivity, in which case the asymptotic behavior of $f$ is controlled by the ballistic continuum, or it may be an isolated zero of $\mathfrak{D}(0,ik)$, which, for nonzero $k$, is a solution of \eqref{dispersionD} with $\omega=0$, and therefore corresponds to a stationary evanescent non-hydrodynamic mode. Singularities located further away from the real $k$ axis describe subleading contributions to the form factor, which decay faster than $e^{-\eta_{\text{max}}r}$.

\begin{figure}[h!]
    \centering
\includegraphics[width=0.43\linewidth]{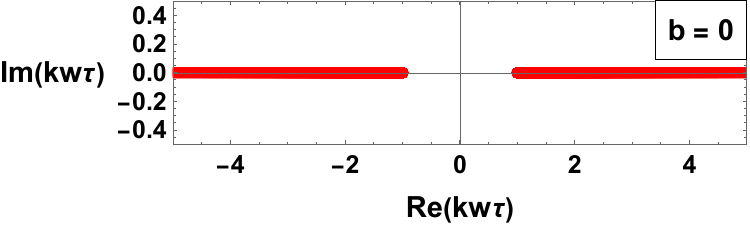}\hspace{0.6cm}\includegraphics[width=0.43\linewidth]{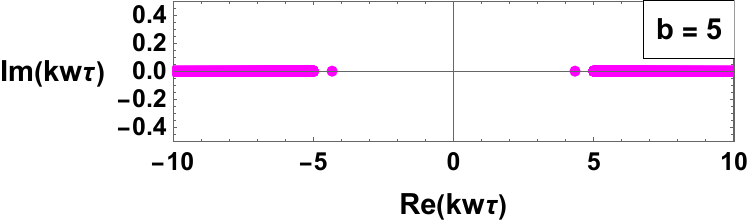}
\includegraphics[width=0.43\linewidth]{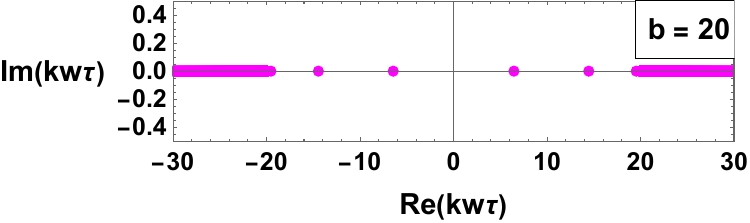}\hspace{0.6cm}\includegraphics[width=0.43\linewidth]{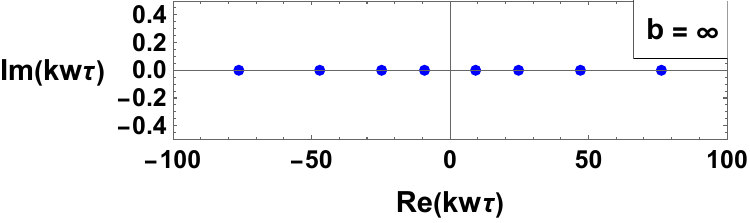}
\includegraphics[width=0.43\linewidth]{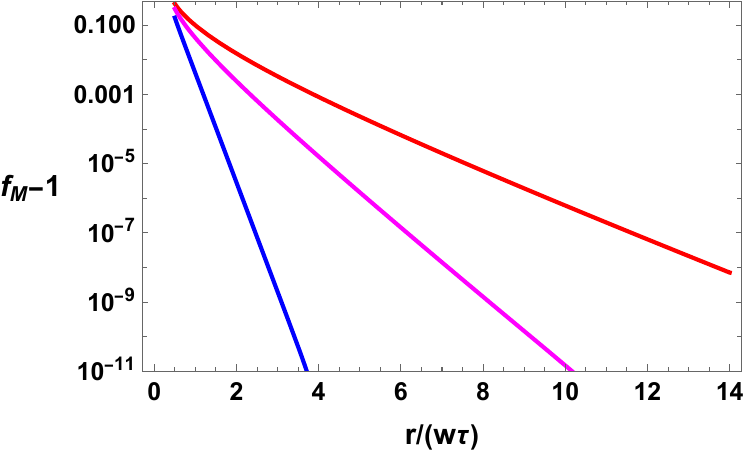}\hspace{0.6cm}\includegraphics[width=0.43\linewidth]{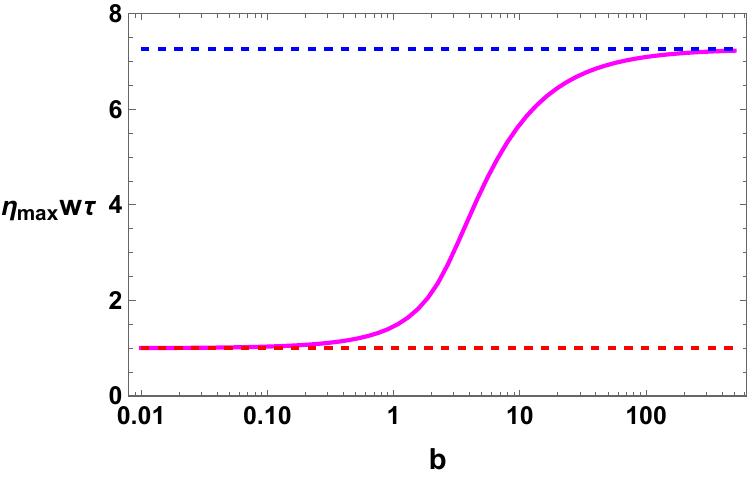}
\caption{Analytic structure controlling the large-$r$ behavior of the mollified form factor. The color coding follows the previous figures: red denotes isotropic scattering, magenta the interpolating scattering kernel \eqref{Fbexponential}, and blue small-angle scattering. \textit{Upper panels}: Singularities of $\mathfrak{D}(0,ik)^{-1}$ in the complex $kw\tau$ plane for $b=0$, $5$, $20$, and $\infty$. In the isotropic limit, the singularities form the ballistic continuum, while increasing $b$ produces isolated singularities that progressively separate from the continuum. In the small-angle limit the continuum is pushed to infinity, leaving only isolated singularities. \textit{Lower left}: Large-$r$ behavior of $f_M(0,r)-1$ for isotropic scattering (red), $b=2$ (magenta), and small-angle scattering (blue), shown on a logarithmic scale. The exponential decay rates are $\eta_{\rm max}w\tau=1$, $2.16$, and $7.26$, respectively. \textit{Lower right}: Theoretical value of $\eta_{\rm max}w\tau$, determined by the singularity of $\mathfrak{D}(0,ik)^{-1}$ closest to the real $k$ axis, as a function of $b$. The dashed red and blue lines denote the isotropic and small-angle limits, $\eta_{\rm max}w\tau=1$ and $7.26$, respectively. The agreement between the two lower panels confirms that the large-distance decay of the form factor is controlled by the analytic structure of the resummed diffusivity in the complex-wavenumber plane.}
\label{fig:logarithmiamo}
\end{figure}

In order to study the singularities of $\mathfrak{D}(0,ik)^{-1}$, we make the following observation. If we stop the chain of equations \eqref{intermediateM} at the second line, we obtain $1/\mathfrak{D}(0,ik)\propto \sigma_1-(ikwa_2)^2[\mathbb{M}^{(2)}]^{-1}_{11}$. Thus, barring accidental cancellations, singularities of $\mathfrak{D}(0,ik)^{-1}$ occur when $\mathbb{M}^{(2)}$ becomes non-invertible. We recall that $\mathbb{M}^{(2)}$ is simply the matrix \eqref{sNnN} with its first row and column removed, so that we can denote it by $\s_2+ikwn_2^x$. Now, since $\|\s_2^{-1}\|\leq\tau$ and $\|n_2^x\|\leq1$, a standard Neumann-series argument immediately gives $\eta_\text{max}\geq(w\tau)^{-1}$, meaning that $f_M-1$ decays at least like $e^{-r/(w\tau)}$, up to, possibly, subexponential factors. Furthermore, $\s_2+ikwn_2^x$ is non-invertible if and only if $-(ikw)^{-1}\in\operatorname{Spectrum}(\s_2^{-1/2}n_2^x\s_2^{-1/2})$,
and since the latter operator is self-adjoint, we conclude that all such singularities are located on the line $ik\in\mathbb{R}$. Physically, this means that the corresponding exponentially suppressed contributions to $f_M-1$ decay without spatial oscillations. Numerically, there is no obstruction to computing the eigenvalues of $\s_2^{-1/2}n_2^x\s_2^{-1/2}$ for large (i.e. order $1000$) harmonic truncations. This allows us to determine $\eta_\text{max}$ for varying $b$ with great accuracy. The result is shown in figure \ref{fig:logarithmiamo}.

\section{Conclusions}
\vspace{-0.4cm}

In this work, we have obtained an exact all-order resummation of the hydrodynamic gradient expansion for monoenergetic particles undergoing elastic scattering in a homogeneous and rotationally invariant medium. The resummed diffusivity $\mathfrak{D}(i\omega,ik)$ takes the form of an infinite continued fraction, whose coefficients are determined by the angular relaxation rates $\{\sigma_\ell\}_{\ell\geq1}$ of the collision operator. While the ordinary diffusion coefficient depends only on $\sigma_1$, progressively higher orders in the gradient expansion probe progressively higher angular harmonics. The fully resummed constitutive relation therefore retains full information about the complete scattering kernel.

We have explored how this microscopic information manifests itself in the analytic structure of $\mathfrak{D}$. Isotropic and small-angle scattering provide two qualitatively different limiting cases: the former exhibits a ballistic branch cut, whereas the latter has only isolated poles. More realistic scattering kernels interpolate continuously between these two regimes, with isolated poles coexisting with a ballistic continuum. We have also shown how the excitation spectrum of the kinetic theory can be reconstructed directly from $\mathfrak{D}$ through the equation $k^2\mathfrak{D}(i\omega,ik)-i\omega=0$. In particular, the hydrodynamic mode may either merge directly into the ballistic continuum or collide with a non-hydrodynamic mode before eventually being absorbed into it, depending on the angular structure of the scattering process.

Beyond dispersion relations, the resummed diffusivity determines the response of the system to external sources. We illustrated this by considering a spatially localized source oscillating at a fixed frequency and introducing a form factor that measures the deviation of the resulting density profile from Fick's law. Although these deviations become rapidly suppressed with distance from the source, their decay retains detailed information about the microscopic scattering kernel. At zero frequency, we showed that the large-distance behavior is controlled by the singularity of $\mathfrak{D}(0,ik)^{-1}$ closest to the real-$k$ axis. Thus, even exponentially small corrections to the hydrodynamic response can provide direct information about the analytic structure of the all-order constitutive relation.

These results illustrate a broader point about resummed hydrodynamics: The transport coefficients appearing at any finite order capture only a limited amount of microscopic information, whereas their all-order resummation organizes the full nonlocal response into a single function \cite{GavassinoDiamond:2026fgz}. Its poles, branch cuts, zeros, and complex-wavenumber singularities encode complementary aspects of the underlying kinetic dynamics. The present model provides an analytically tractable setting in which this connection can be followed explicitly from the microscopic scattering integral to observable departures from Fick's law.

\vspace{-0.4cm}
\section*{Acknowledgements}
\vspace{-0.3cm}

This work is supported by a MERAC Foundation prize grant,  an Isaac Newton Trust Grant, and funding from the Cambridge Centre for Theoretical Cosmology.

\appendix
\vspace{-0.4cm}
\section{Closed-form expression for the isotropic scattering diffusivity}\label{AAA}
\vspace{-0.3cm}

\begin{figure}[h!]
    \centering
\includegraphics[width=0.46\linewidth]{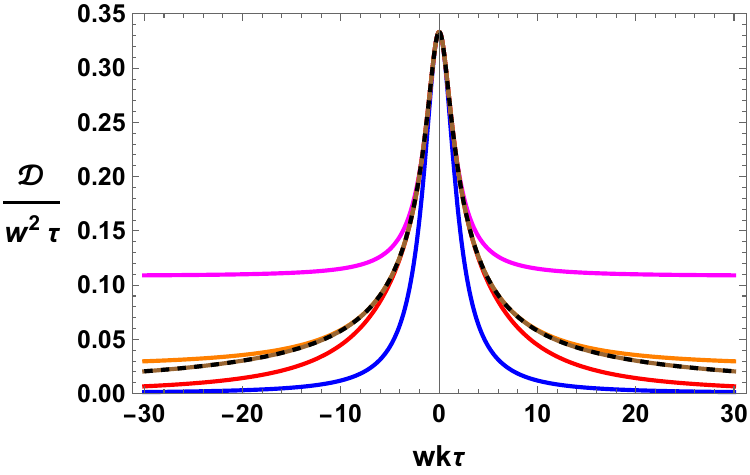}\hspace{0.6cm}\includegraphics[width=0.46\linewidth]{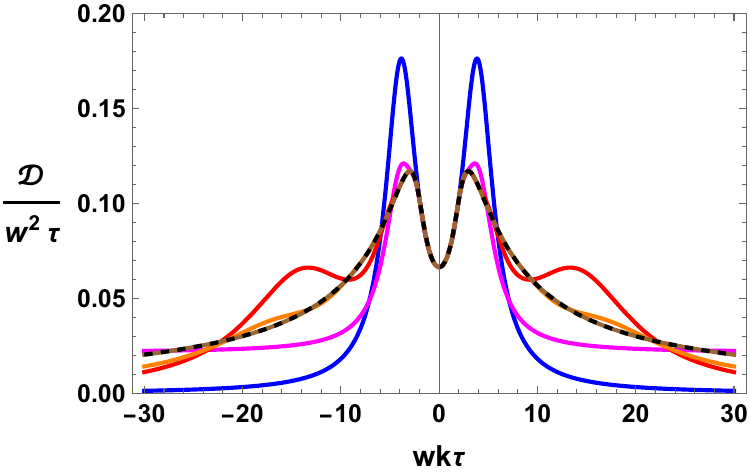}
\caption{Convergence test of the truncated continued fraction \eqref{Diomik} for isotropic scattering, with $\omega=0$ (left panel) and $\omega=2/\tau$ (right panel). The continuous curves correspond to truncations at $\ell=2$ (blue), $5$ (magenta), $10$ (red), $20$ (orange), and $100$ (brown), while the black dashed curve is the exact result \eqref{exactD}. As the truncation order alternates between even and odd values, $\mathfrak{D}_\ell$ oscillates around the exact result and eventually converges to it. The value of $\ell$ required to achieve convergence increases with $|k|$, reflecting the nonuniformity of the convergence in $k$ and, ultimately, the fact that the limits $\ell\to\infty$ and $|k|\to\infty$ do not commute.}
\label{fig:Convergencetest}
\end{figure}

For isotropic scattering, we have $\s_N=\tau^{-1}$. Hence, \eqref{Dwa1} reduces to 
\begin{equation}\label{diffusivitymatrixlolol}
\mathfrak{D}(i\omega,ik)=\dfrac{w^2\tau}{3(1-i\omega\tau)} (Y^0_1,[1+r n^x_N]^{-1}Y^0_1) \, , \qquad \qquad \text{with } r=\dfrac{ikw\tau}{1-i\omega \tau}\, .
\end{equation}
Now, introducing the state $\Phi=[1+r n^x_N]^{-1}Y^0_1 \in (Y_0^0)^\perp$, we have $[1+r \PN n^x]\Phi=Y^0_1$, or, equivalently,
\begin{equation}\label{PhiIsthis}
\Phi=\dfrac{Y^0_1+r Y_0^0 (Y_0^0,n^x\Phi)}{1+rn^x} \, .
\end{equation}
Taking the inner product with $Y_0^0$, and recalling that $\Phi$ is orthogonal to it, we obtain $(Y_0^0,n^x\Phi)$, which, inserted back into \eqref{PhiIsthis}, gives
\begin{equation}\label{PhiIsthis2}
\Phi=
\dfrac{Y_1^0}{1+rn^x}
-\dfrac{Y_0^0}{1+rn^x}
\dfrac{\left(Y_0^0,\dfrac{Y_1^0}{1+rn^x}\right)}
{\left(Y_0^0,\dfrac{Y_0^0}{1+rn^x}\right)}
\equiv [1+r n_N^x]^{-1}Y_1^0 \, .
\end{equation}
Inserting this expression into \eqref{diffusivitymatrixlolol}, and evaluating the inner products explicitly, we finally obtain
\begin{equation}\label{exactD}
\mathfrak{D}(i\omega,ik)=
\dfrac{w^2\tau(1{-}i\omega\tau)}{(ikw\tau)^2}
\left[
1-\dfrac{\frac{ikw\tau}{1-i\omega\tau}}
{\operatorname{arctanh}\left(\frac{ikw\tau}{1-i\omega\tau}\right)}
\right] \, .
\end{equation}
In Fig. \ref{fig:Convergencetest}, we show the convergence of the continued fraction to this exact result.

\section{Large-$k$ behavior for unbounded scattering operators}\label{BBB}

Consider a rotationally invariant scattering operator with positive angular relaxation rates $\{\sigma_\ell\}_{\ell\geq 1}$. At $\omega=0$, equation \eqref{Diomik} can be rearranged as follows:
\begin{equation}
\mathfrak{D}(0,ik)=\dfrac{w}{k}
\dfrac{a_1^2}{
\dfrac{\sigma_1}{kw}
+\dfrac{a_2^2}{
\dfrac{\sigma_2}{kw}
+\dfrac{a_3^2}{
\dfrac{\sigma_3}{kw}
+\dfrac{a_4^2}{
\dfrac{\sigma_4}{kw}
+\dfrac{a_5^2}{
\dfrac{\sigma_5}{kw}+\ddots}}}}} \, .
\end{equation}
Now, fix $L\gg 1$ odd, and call $T_L$ the tail of the continued fraction that starts at $a_L^2$. For sufficiently large $k$, all terms $\sigma_\ell/(kw)$ with $\ell\leq 2L$ are much smaller than one. Hence, we have the following limiting behavior:
\begin{equation}
\mathfrak{D}\simeq \dfrac{w}{k}\,
\underbrace{
\dfrac{a_1^2}{a_2^2}
\dfrac{a_3^2}{a_4^2}
\dfrac{a_5^2}{a_6^2}
\cdots
\dfrac{a_{L-2}^2}{a_{L-1}^2}
}_{\text{Approaches }4/\pi}
\,T_L \, .
\end{equation}
Our goal is now to determine $T_L$.

Since $a_\ell\to 1/2$ for large $\ell$, the remaining tail follows the recursion relation
\begin{equation}\label{recurrenziamo}
T_\ell=\dfrac{1}{4(\varepsilon_\ell+T_{\ell+1})} \, ,
\qquad\qquad
L\leq \ell\leq 2L \, ,
\end{equation}
where $\varepsilon_\ell=\sigma_\ell/(kw)$ are very small numbers throughout this range. We immediately see that $1/2$ is a fixed point of this recursion relation in the limit $\varepsilon_\ell\to0$. Moreover, defining $\Delta_\ell=T_\ell-1/2$, and applying the recursion relation twice, we obtain
\begin{equation}\label{B4}
\dfrac{\Delta_{\ell-1}}{\Delta_{\ell+1}}
=
1-\dfrac{\varepsilon_\ell}{\Delta_{\ell+1}}
\left[
\dfrac{(1{+}2\Delta_{\ell+1})^2\varepsilon_{\ell-1}}
{\varepsilon_\ell}-1
\right]
+\mathcal{O}(\varepsilon^2) \, .
\end{equation}

In the case of isotropic scattering, we have $\varepsilon_\ell=\text{const}\equiv\varepsilon$, and the above expression reduces to
\begin{equation}\label{elgringo}
\dfrac{\Delta_{\ell-1}}{\Delta_{\ell+1}}
=
1-4\varepsilon\left(1+\Delta_{\ell+1}\right)
+\mathcal{O}(\varepsilon^2) \, .
\end{equation}
Thus, as we apply the recursion relation repeatedly, $\Delta_\ell$ is driven toward zero. More precisely, starting from an arbitrary value of $T_{2L}$, by the time we reach $T_L$ the distance from $1/2$ has been reduced by a factor of order $e^{-2\varepsilon L}$. Taking $L$ very large while holding $k$ (and hence $\varepsilon$) fixed, this exponential tends to zero, so that $T_L\to1/2$. Since the prefactor in the previous product approaches $4/\pi$, we recover the asymptotic behavior $\mathfrak{D}(0,ik)\simeq 2w/(\pi k)$, as expected. 

Consider now the small-angle scattering limit, for which $\varepsilon_\ell\propto\ell(\ell+1)$. We still have $\varepsilon_{\ell-1}/\varepsilon_\ell\to1$ for large $\ell$. Hence, the attraction law \eqref{elgringo} continues to hold locally, but now with an $\ell$-dependent $\varepsilon$. Since $\varepsilon_\ell\geq\varepsilon_L$ for $\ell\geq L$, the ratio $|\Delta_L/\Delta_{2L}|$ is suppressed by at least a factor of order $e^{-2\varepsilon_L L}$.
We now wish to send $L\to\infty$. In this case, however, we cannot hold $k$ fixed, because we require $\varepsilon_\ell$ to remain small all the way to $2L$. We therefore take the large-$L$ and large-$k$ limits simultaneously, fixing $\varepsilon_{2L}=\varepsilon\ll1$.
For small-angle scattering, $\varepsilon_L/\varepsilon_{2L}\to 1/4$, so that
\begin{equation}
e^{-2\varepsilon_L L}
=
e^{-2\frac{\varepsilon_L}{\varepsilon_{2L}}\varepsilon L}
\simeq
e^{-\varepsilon L/2}
\longrightarrow0 \, .
\end{equation}
Thus, $T_L\to1/2$ also in the small-angle scattering limit, and we recover again $\mathfrak{D}(0,ik)\simeq 2w/(\pi k)$.

We conclude that, although the eigenvalues of the small-angle scattering operator grow without bound, they do not grow rapidly enough to prevent the resummed diffusivity from approaching the free-streaming result at large $k$. This is particularly striking because the same unbounded growth prevents the existence of a continuous ballistic branch \cite[Theorem 5]{GavassinoLorentzian:2026seq}. Thus, the emergence of free-streaming asymptotics in the constitutive response does not require the existence of ballistic excitations in the spectrum.

Finally, we stress that, for different growth laws of the eigenvalues, our argument can fail in two ways. First, the square bracket in \eqref{B4} may become negative, in which case the tails are no longer attracted toward $1/2$. Second, the eigenvalues may vary in such a way that, even if $1/2$ remains an attractive fixed point, the accumulated suppression does not tend to zero as $L\to\infty$, owing to the need to progressively increase $k$ in order for $\varepsilon_{2L}$ to remain small.

\section{Closed-form expression for the resummed diffusivity in the first-order anisotropic approximation}\label{CCC}

Our starting point is the continued fraction
\begin{equation}\label{Continuedquasiisotrop}
\mathfrak{D}(i\omega,ik)= \dfrac{w^2 a_1^2}{\tau^{-1}{-}i\omega-\dfrac{(ikwa_2)^2}{\Tilde{\tau}^{-1}{-}i\omega -\dfrac{(ikwa_3)^2}{\Tilde{\tau}^{-1}{-}i\omega-\dfrac{(ikwa_4)^2}{\Tilde{\tau}^{-1}{-}i\omega \ddots}} } } \, ,
\end{equation}
where we have introduced $\Tilde{\tau}=\tau (1-\frac{b}{3})$. Now, we recall that the isotropic scattering solution \eqref{exactD} also admits a continued fraction representation, which results in the identity (replacing $\tau$ with $\Tilde{\tau}$)
\begin{equation}
\dfrac{\Tilde{\tau}(1{-}i\omega\Tilde{\tau})}{(ikw\Tilde{\tau})^2}
\left[
1-\dfrac{\frac{ikw\Tilde{\tau}}{1-i\omega\Tilde{\tau}}}
{\operatorname{arctanh}\left(\frac{ikw\Tilde{\tau}}{1-i\omega\Tilde{\tau}}\right)}
\right]=\dfrac{a_1^2}{\Tilde{\tau}^{-1}{-}i\omega-\dfrac{(ikwa_2)^2}{\Tilde{\tau}^{-1}{-}i\omega -\dfrac{(ikwa_3)^2}{\Tilde{\tau}^{-1}{-}i\omega-\dfrac{(ikwa_4)^2}{\Tilde{\tau}^{-1}{-}i\omega \ddots}} } } \, ,
\end{equation}
which can be rearranged as follows
\begin{equation}
\left(\dfrac{\Tilde{\tau}(1{-}i\omega\Tilde{\tau})}{a_1^2(ikw\Tilde{\tau})^2}
\left[
1-\dfrac{\frac{ikw\Tilde{\tau}}{1-i\omega\Tilde{\tau}}}
{\operatorname{arctanh}\left(\frac{ikw\Tilde{\tau}}{1-i\omega\Tilde{\tau}}\right)}
\right]\right)^{-1}-\Tilde{\tau}^{-1}={-}i\omega-\dfrac{(ikwa_2)^2}{\Tilde{\tau}^{-1}{-}i\omega -\dfrac{(ikwa_3)^2}{\Tilde{\tau}^{-1}{-}i\omega-\dfrac{(ikwa_4)^2}{\Tilde{\tau}^{-1}{-}i\omega \ddots}} } \, .
\end{equation}
We see that the right-hand side coincides with the tail of \eqref{Continuedquasiisotrop}, so we immediately obtain
\begin{equation}
\mathfrak{D}(i\omega,ik)=\dfrac{\Tilde{\tau} w^2}{\dfrac{(ikw\Tilde{\tau})^2}{1{-}i\omega\Tilde{\tau}}
\left[
1-\dfrac{\frac{ikw\Tilde{\tau}}{1-i\omega\Tilde{\tau}}}
{\operatorname{arctanh}\left(\frac{ikw\Tilde{\tau}}{1-i\omega\Tilde{\tau}}\right)}
\right]^{-1}-b} \, .
\end{equation}

\bibliography{Biblio}

\label{lastpage}
\end{document}